\documentclass[%
reprint,
superscriptaddress,
showpacs,
preprintnumbers,
 amsmath,amssymb,
 aps,
prc,
twocolumn,
floats
]{revtex4-1}

\usepackage{graphicx}
\usepackage{dcolumn}
\usepackage{bm}
\usepackage{epstopdf}
\usepackage{subfigure}

\begin{document}

\title{Interpretation of the large-deformation
high spin bands in selected $A=158-168$ nuclei}

\author{A.~Kardan}
\affiliation{Division of Mathematical Physics, LTH, Lund University, Post Office Box 118, S-22100 Lund, Sweden}
\affiliation{Physics Department, Faculty of Science, Ferdowsi University of Mashhad, P.O. Box 91775-1436, Mashhad, Iran}
\author{I.~Ragnarsson}
\affiliation{Division of Mathematical Physics, LTH, Lund University, Post Office Box 118, S-22100 Lund, Sweden}
\author{H.~Miri-Hakimabad}
\author{L.~Rafat-Motevali}
\affiliation{Physics Department, Faculty of Science, Ferdowsi University of Mashhad, P.O. Box 91775-1436, Mashhad, Iran}

\date{\today}

\begin{abstract}
The high-spin rotational bands in $^{168}$Hf and the triaxial bands in Lu
nuclei are analyzed using the configuration-constrained
Cranked Nilsson-Strutinsky (CNS) model. Special attention is given to the
up-sloping extruder orbitals. The relative alignment between the bands
which appear to correspond to triaxial shape
is also considered, including the yrast ultra-high spin band in $^{158}$Er.
This comparison suggests that the latter band is formed from rotation around the
intermediate axis.
In addition, the standard approximations of the CNS approach are investigated, indicating that
the errors which are introduced by the 
neglect of off-shell matrix elements and the cut-off at 9 oscillator shells (${\cal N}_{max}=8$) 
are essentially negligible compared to other uncertainties. On the other
hand, the full inclusion of the hexadecapole degree of freedom is more
significant; for example it
leads to a decrease of the total energy of $\sim 500$ keV in the TSD region of
$^{168}$Hf. 
\end{abstract}


\maketitle

\section{Introduction}
\label{introd}

The high-spin structure of deformed nuclei shows a variety of interesting
phenomena caused by the interplay between collective and single-particle
excitations.
The region of nuclei 
with $Z\sim72$ and $N\sim94$ is particularly fascinating. 
Potential energy surface (PES) calculations,
predict that these nuclei constitute a new region of exotic 
shapes~\cite{IR,SA,Naz91}
coexisting with normal prolate deformation
($\varepsilon_{2}\sim0.23$). At high spins these nuclei
may assume stable triaxial superdeformed (TSD) shapes characterized
by different moments of inertia for each of the principal axes.
These TSD minima, with
deformation parameters $(\varepsilon_{2},\gamma)\sim(0.4,\pm20^{\circ})$,
are caused by large single-particle shell gaps associated
with proton numbers $Z=71$ and $72$, and neutron numbers $N=94$
and $96$~\cite{HS,BR}. Experimentally, such rotational bands have been
reported in Lu ($Z=71$) isotopes~\cite{SO,GS,HA}. 

An extensive search for
TSD bands in Hf ($Z=72$) isotopes has also been carried out, and
a number of strongly deformed bands have been observed in
$^{170-175}$Hf~\cite{AN,YZ,MD,DH,DS},
where bands in $^{170}$Hf~\cite{AN} and $^{174}$Hf~\cite{MD}
have been tentatively assigned as triaxial. On the other
hand, the predicted TSD bands in $^{164}$Hf and $^{166}$Hf have
not been discovered. Indeed, according to the 
analysis in Ref.~\cite{YZ}, all observed strongly deformed bands in
$^{170-175}$Hf are most likely near prolate falling into two groups
corresponding enhanced deformation (ED) shapes 
(deformations enhanced with respect to the normal
deformed nuclear shapes) and superdeformed (SD) shapes.
The ED bands with $\varepsilon_{2}\sim0.3$ are built
on the proton $i_{13/2}h_{9/2}$ configuration while the 
SD bands involve the $\pi i_{13/2}$ (proton) and $\nu j_{15/2}$ (neutron)
orbitals.
On the other hand,
a high-spin band has been observed in $^{168}$Hf~\cite{HA2,RY,RY2}
which appears to correspond to triaxial shape with a deformation which
is considerably larger than that of the TSD bands in $^{161-167}$Lu.


The high-spin bands which have attracted most interest recently are
however the so-called ultrahigh-spin bands which bypass the 
band-terminating states in $^{157-158}$Er~\cite{Pau07}
and neighboring nuclei~\cite{Agu08,Tea08,Oll09}.
These bands were first assumed to have a triaxial deformation similar
to that of the TSD bands in  Lu nuclei but recent lifetime measurements~\cite{Wan11}
show that they are more collective and they are suggested to correspond
to either a larger triaxial deformation or possibly a similar deformation as
the Lu TSD bands but with rotation around the intermediate axis ($\gamma <
0$). In a recent study \cite{Shi12}, it was concluded that these bands must
correspond to a larger triaxial deformation because the $\gamma < 0$
minimum appears to be a saddle point if the rotation axis is allowed
to change direction. In any case,
it has turned out to be difficult to find a consistent
interpretation within the standard CNS approach~\cite{AA,TB,BGC}.
This is one reason why
it appears important to investigate if, within the CNS approach, 
it is possible to get a
consistent interpretation of the unique large deformation TSD
bands which have been observed 
in $^{168}$Hf. In this context, we will also demonstrate that 
the smaller deformation TSD
bands in Lu isotopes appears to get a ready interpretation
in the CNS formalism, see also~\cite{Rag08}. 

Partly because of the large deformation
of the TSD band in $^{168}$Hf, some approximations of the CNS 
approach become somewhat questionable. Therefore, we have made some modifications in the formalism making it 
possible to investigate the importance to include more
oscillator shells in the basis and to account for all
matrix elements coupling the different ${\cal N}$-shells 
of the
harmonic oscillator basis. Most important however is
that, for the first time to our knowledge, a complete
minimization in the three hexadecapole degrees of freedom
has been carried out at a large triaxial deformation.



The motivation for the present work is to study high-spin rotational bands in $^{168}$Hf and
investigate their properties in order to understand their nature. 
As a background, we will consider the TSD bands in the Lu isotopes.
The Er bands have already been analyzed in Refs. \cite{Pau07,Wan11} but
we will conclude with some additional comments. 
We do the calculations within 
the framework of the configuration-constrained
Cranked Nilsson-Strutinsky (CNS) model~\cite{AA,TB,BGC} and another
motivation is to test and develop this formalism. 
The model and standard approximations are explained in sect.~\ref{model}. A brief description 
of the structure of the observed TSD bands 
in Lu isotopes using the CNS formalism is presented in sect.~\ref{Lu}. 
Standard approximations of the CNS formalism 
are tested in sect.~\ref{off} while a complete
minimization
in the hexadecapole space is carried out in sect.~\ref{min}.
The reference energy which is often subtracted when presenting
nuclear high-spin bands is discussed in sect.~\ref{ref}. 
Then we
study the experimental and theoretical high-spin bands in $^{168}$Hf in sects.~\ref{obs} and 
~\ref{calc}. In sect.~\ref{comp},
we compare these theoretical and the experimental bands and find out which
theoretical bands correspond 
to band 1, band 3, the ED band and the TSD1 and TSD2 bands of $^{168}$Hf.
Finally, we
present some new points of view for the yrast ultrahigh-spin $^{158}$Er band in sect.~\ref{Er}. 

\section{The standard CNS formalism}
\label{model}


In the configuration-dependent
Cranked Nilsson-Strutinsky (CNS) model~\cite{TB,AA,BGC},
the nucleons are moving independently of each other in a deformed
and rotating mean-field generated by the nucleons themselves. The rotation or the effect of the rotation is treated as an external potential. The
mean-field Hamiltonian used to describe a nucleon in the rotating nucleus is the cranked modified oscillator Hamiltonian~\cite{TB} 
\begin{eqnarray}
H &=& h_{HO}(\varepsilon_{2},\gamma)-\kappa\hbar\omega_{0}(2\ell_{t}\cdot
s+\mu(\ell_{t}^{2}- \langle \ell_{t}^{2} \rangle_{N})) \nonumber\\
&+& V_{4}(\varepsilon_{4},\gamma)-\omega j_{x}\label{eq:1}
\end{eqnarray}
In this Hamiltonian, the cranking term $\omega j_{x}$ is introduced to make
the deformed potential rotate uniformly around a principal axis
with the angular velocity $\omega$. The index
$t$ in the orbital angular momentum operator $\ell_{t}$, denotes that
it is defined in stretched coordinates~\cite{SN,book}.
For $^{168}$Hf, standard values~\cite{TB} are used for
the single-particle parameters 
$\kappa$ and $\mu$, which determine 
the strength of the $\ell_{t}\cdot s$ and $\ell_{t}^{2}$ terms,
while $A=150$ parameters \cite{TB90} are used for the $^{161-167}$Lu and $^{158}$Er.
This is motivated by the fact that the $A=150$ parameters have been fitted
for nuclei with $N \approx 90$, while standard parameters should be more 
appropriate for the
well-deformed nuclei in the middle of the rare-earth region.

In Eq.~(\ref{eq:1}),  
$h_{HO}(\varepsilon_2,\gamma)$ is an anisotropic harmonic-oscillator Hamiltonian:
\begin{eqnarray}
h_{HO}(\varepsilon_2,\gamma)=\frac{p^{2}}{2m}+\frac{1}{2}m(\omega_{x}^{2}x^{2}+\omega_{y}^{2}y^{2}+\omega_{z}^{2}z^{2})\label{eq:ho}
\end{eqnarray}
The relation between the oscillator frequencies and $\varepsilon_2$, $\gamma$ is:
\begin{eqnarray}
\omega_{x} & = & \omega_{0}(\varepsilon_2,\gamma)\:(1-\frac{2}{3}\varepsilon_2\,\cos(\gamma+\frac{2\pi}{3}))
\nonumber \\
\omega_{y} & = & \omega_{0}(\varepsilon_2,\gamma)\:(1-\frac{2}{3}\varepsilon_2\,\cos(\gamma-\frac{2\pi}{3}))\label{eq:omega} \\
\omega_{z} & = & \omega_{0}(\varepsilon_2,\gamma)\:(1-\frac{2}{3}\varepsilon_2\,\cos\gamma)\nonumber
\end{eqnarray}
The deformation dependence of $\omega_{0}(\varepsilon_2,\gamma)$ is determined from volume conservation of the equipotential surfaces. 

The total energy is obtained using the shell correction method. Thus the shell energy, $E_{sh}$, is calculated using the Strutinsky procedure ~\cite{VS,GA}
 and the total energy is defined as the sum of the shell energy and the rotating liquid drop energy~\cite{BGC,GA}, $E_{rld}$,
\begin{equation}
 E_{tot}(I)=E_{sh}(I)+E_{rld}(I).\label{eq:sh}
\end{equation}

This renormalization ensures that the total nuclear energy is correct on the
average. 
The Lublin Strasbourg drop model~\cite{KP2} is used for the static liquid drop
energy with the rigid-body moment of inertia calculated with a radius
parameter $r_{0}=1.16$ fm and a diffuseness parameter $a=0.6$ fm~\cite{BGC}. 
Finally, minimizing the total energy for a given angular momentum
with respect to deformation gives the equilibrium shape and corresponding energy.
Plots of the minimized total energy versus spin $I$ are frequently used in the 
description of high-spin properties of rotating nuclei. To present considerably 
more detailed information about individual and relative properties of the rotational 
bands, the excitation energy is plotted relative to a reference energy. 
Note that the same reference energy is utilized for all theoretical and experimental energies
 in a nucleus.

 Eq.~(\ref{eq:1}) represents the rotating modified oscillator Hamiltonian
in terms of the quadrupole, $\varepsilon_{2}$, non-axial, $\gamma$,
and the hexadecapole, $\varepsilon_{4}$, deformation parameters.
The dependence of the Hamiltonian on the hexadecapole deformation
is written as:
\begin{eqnarray}
V_{4} & = &
2\hbar\omega_{0}\rho^{2}\left[\varepsilon_{40}Y_{4}^{0}\left(\theta_{t},\varphi_{t}\right) \right.
\label{eq:2}
\\
& + & \varepsilon_{42}\left(Y_{4}^{2}\left(\theta_{t},\varphi_{t}\right)+Y_{4}^{-2}\left(\theta_{t},
\varphi_{t}\right)\right)\nonumber\\
& + & \left.
\varepsilon_{44}\left(Y_{4}^{4}\left(\theta_{t},\varphi_{t}\right)+Y_{4}^{-4}\left(\theta_{t},\varphi_{t}\right)\right)\right],
\nonumber
\end{eqnarray}
with~\cite{TB,SR}
\begin{eqnarray}
\varepsilon_{40} & = & \varepsilon_{4}\frac{1}{6}(5\cos^{2}\gamma+1)\nonumber \\
\varepsilon_{42} & = & -\varepsilon_{4}\frac{1}{12}\sqrt{30}\sin2\gamma\label{eq:3} \\
\varepsilon_{44} & = & \varepsilon_{4}\frac{1}{12}\sqrt{70}\sin^{2}\gamma,\nonumber
\end{eqnarray}
where $\theta_{t}$ and $\varphi_{t}$ are the polar and
azimuthal angles in stretched coordinates and $\rho$ is the radius
in stretched coordinates. The $\gamma$ dependence in Eq.~(\ref{eq:3}) is
introduced in such a way that the axial symmetry is preserved
when $\gamma=-120^{\circ},\,-60^{\circ},\,0^{\circ}$
or $60^{\circ}$. 
All ellipsoidal shapes can be described within a $60^{\circ}$ degree sector,
but the rotation occurs around the shortest, the intermediate and the
longest principal axis for $\gamma = [0^{\circ},\,60^{\circ}]$, 
$\gamma = [0^{\circ},\,-60^{\circ}]$ and $\gamma =
[-60^{\circ},\,-120^{\circ}]$, respectively.


Because the $\varepsilon_{4i}$ parameters
depend on one parameter $\varepsilon_{4}$, there is
only one hexadecapole degree of freedom. In a standard calculation,
the total energy is minimized varying three parameters: 
two quadrupole parameters, $\varepsilon_2$ and $\gamma$, and one hexadecapole parameter, $\varepsilon_4$~\cite{TB}. 
The choice of the deformation space to be used in a calculation is
important. Recently, some studies concentrating on the
role of different multipoles on the fission barrier heights have considered
more general hexadecapole deformations~\cite{JD,ASt,ASo}. 

The rotating basis $| n_{x}n_{2}n_{3}\Sigma \rangle$ can be utilized
to diagonalize the Hamiltonian matrix and to find eigenfunctions of
Eq.~(\ref{eq:1})~\cite{TB}. Since the couplings of $j_{x}$ are fully accounted
for in the rotating basis, the only terms in Eq.~(\ref{eq:1}) which couple between
basis states of different ${\cal N}_{rot}=n_{x}+n_{2}+n_{3}$ are the hexadecapole
deformation potential $V_{4}$, and the $\ell_{t}\cdot s$ and
$\ell_{t}^{2}$ terms. The off-shell matrix elements of the latter
terms are small for reasonable rotational frequencies. The importance
of the off-shell matrix elements of the $V_{4}$ term depend on
the deformation region where hexadecapole deformations generally become more
important with increasing quadrupole deformation. For small $\varepsilon_{4}$ values it
thus seems reasonable to neglect all those matrix elements which are
off-shell in the rotating basis and keep ${\cal N}_{rot}$ as a preserved
quantum number. The important advantage of the rotating basis is that
${\cal N}_{rot}$ (generally referred to as ${\cal N}$ below) 
can be treated as an exact quantum number making it possible to
fix configurations in great detail. It seems that this is the most important
feature explaining the success of the CNS approach; especially the possibility
to follow e.g. terminating bands in spin regions where they are not yrast.

The diagonalization of the Hamiltonian, Eq.~(\ref{eq:1}), gives the eigenvalues 
${e_i}^\omega$, which are referred to as the single-particle energies
in the rotating frame or the Routhians. Subsequently, it is straightforward
to calculate different expectation
values like $\langle j_x \rangle$ and $\langle j^2 \rangle$.
The diagonalization of the Hamiltonian is performed with a
cut-off in the single-particle basis which may lead to
errors in the results. The original CNS codes were written with only
9 oscillator shells (${\cal N}_{max}=8$) in the basis and this is the maximum number of shells
which has been used in all subsequent CNS calculations, e.g.~\cite{Rag93,AA,Pau07}.
It seems important to test these approximations, i.e.
the neglect the off-shell hexadecapole matrix elements and the
cut-off in the rotating single-particle basis.

In the present calculations, pairing correlations are neglected, although, it is quite evident that the pairing
field is essential for the description of atomic nuclei~\cite{book2}. This is seen for example from  the observed energy gaps
and the suppression of the moments of inertia in rotating nuclei.
However, it appears that the most of the properties of nuclei at high spins are rather insensitive to the pairing field. 
For example, rotational bands have been studied by the Cranked Nilsson-Strutinsky approach~\cite{TB,AA,BGC},
the Cranked relativistic mean field theory~\cite{Koe89,Ring93,Afan96} not including pair correlations and the Cranked
 relativistic Hartree-Bogoliubov formalism~\cite{Afan99,Afan00,Vre05} including pair correlations. These studies
show that in high-spin regime, calculations without pairing describe the data accurately.
In view of this, it is often advantageous to carry out calculations in an unpaired formalism because of the
more transparent description and, for the present CNS calculations, the unique possibilities to
fix configurations, making it possible to follow for example the drastic shape changes in terminating bands~\cite{AA,Afan05}.

In order to evaluate the importance of the pairing energy in the odd-odd $^{76}$Rb nucleus, rotational bands
have been studied by the Cranked Nilsson-Strutinsky-Bogoliubov (CNSB)
formalism presented in Ref.~\cite{Car08} with particle number projection and with energy
 minimization not only in the shape degrees of freedom, $\varepsilon_2$, $\gamma$ and $\varepsilon_4$
 but also in the pairing degrees of freedom, $\varDelta$ and $\lambda$ and 
have been compared with the predictions of the CNS model~\cite{Wad11}. In these calculations, the contributions
from pairing are found to be small at low spin values and they decrease with increasing spin. The pairing
 energies do not change the general structure which means that, for example, the potential energy surfaces with
 pairing included are found to be very similar to those in the CNS formalism.

The outcome from CNS and CNSB calculations have also been compared in $^{161}$Lu~\cite{Rag10,Ma}.
It turns out that for $I>30$, the inclusion of pairing will correspond to a small renormalization of the moment of
inertia but it does not affect the general structure of the yrast line, band crossings etc. Especially, the
terminating states for $I\sim50$ are essentially unaffected by pairing correlations. With this in mind, we
will analyze the high-spin states of Lu isotopes and $^{168}$Hf in the unpaired CNS formalism where our main interest are those
configurations which cannot be isolated in present formalisms with pairing included. 

For $A=158-168$ nuclei, it is convenient to label the
configurations by the dominant amplitudes of the occupied orbitals and holes
relative to the $^{146}$Gd $(Z=64, N=82)$ closed core; that is,
\begin{eqnarray}
 &&\pi(h_{11/2})^{p_1}(h_{9/2}f_{7/2})^{p_2}(i_{13/2})^{p_3}\nonumber\\
&&\nu({\cal N}=4)^{-n_1}(h_{11/2})^{-n_2}(i_{13/2})^{n_3}(i_{11/2},g_{9/2})^{n_4}(j_{15/2})^{n_5},\nonumber
\end{eqnarray}
where the number of the ${\cal N}=4$ protons and $h_{9/2},f_{7/2}$
neutrons is determined from the total number of protons and
neutrons in a nucleus.
We will often use the shorthand 
notation (where the numbers in parentheses are omitted when they are
equal to zero), 
\[
[p_{1}(p_{2}p_3), (n_{1}n_{2})n_{3}(n_{4}n_{5})].
\]
Note however that this is only for the purpose of labeling the
configurations; in the numerical calculations no core is introduced and
all or most of the couplings between $j$-shells are accounted for according
to the different approximation schemes. 

\section{TSD bands in Lu isotopes}
\label{Lu}
The TSD bands in Lu nuclei are characterized by an odd $i_{13/2}$ proton which
plays an important role for the wobbling excitation \cite{SO}. Apart from
this, the occupied orbitals in these bands have not been given much
attendance. An exception is Ref. \cite{BR}, where the
single-particle orbitals and the corresponding shell gaps at TSD deformation
were discussed. Here we will try to demonstrate the filling
of the orbitals in the lowest TSD bands, indicating the contribution of the,
specific orbitals which become occupied when the number of neutrons increases.
This is analogous to previous classifications of the superdeformed bands
in the $A=150$ region \cite{Rag93,Haa93,Afa98}. A preliminary report of
the present classification was given at the NS2008 conference \cite{Rag08}.
 
\begin{figure}[ht]
\centering\includegraphics[clip=true,width=0.47\textwidth,angle=0]{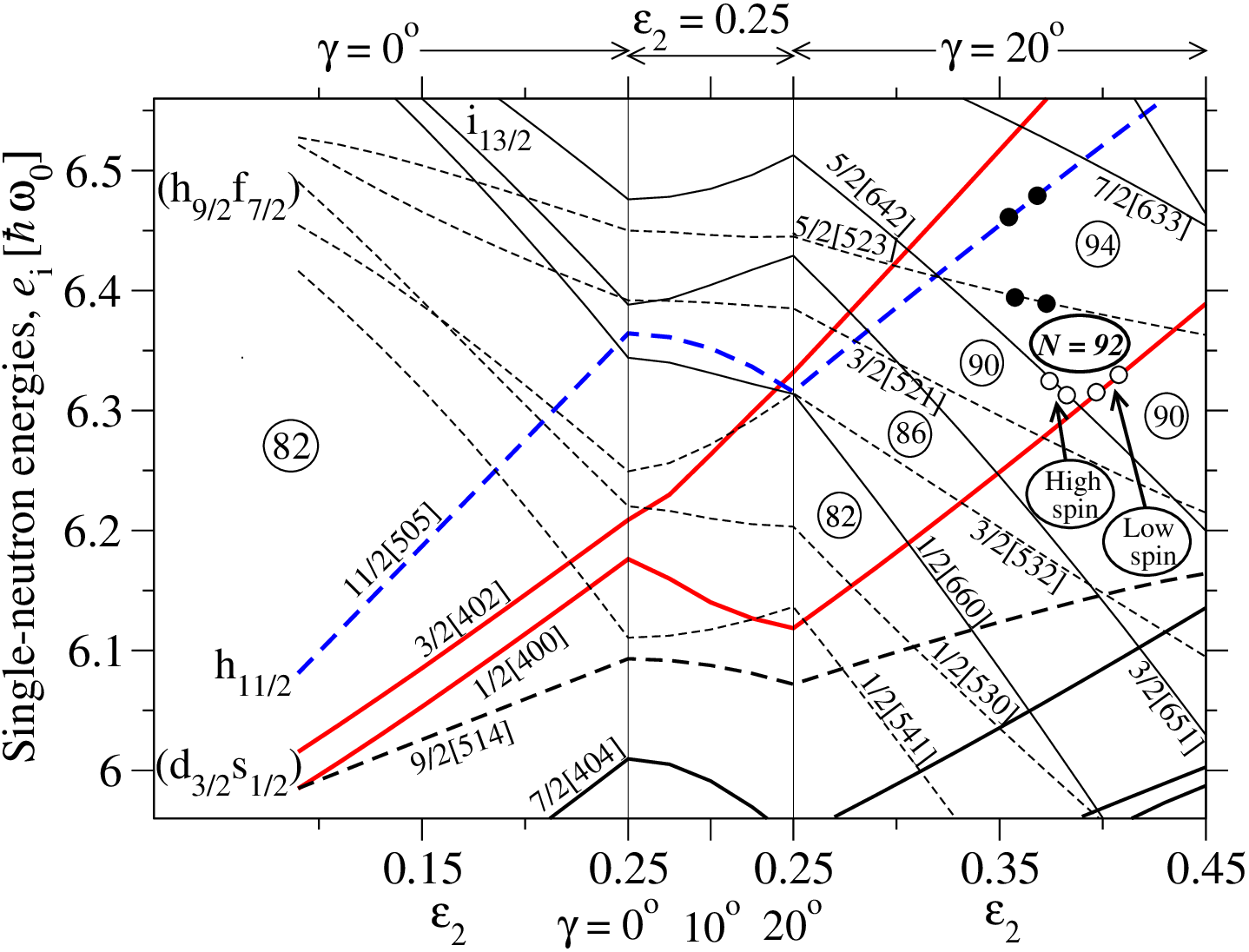}
\caption{\label{spn} {\small(Color online) The single-neutron orbitals drawn along a path in the 
$(\varepsilon_2, \gamma)$-plane in order to clarify the origin of valence
orbitals in the TSD minimum of $A=160-170$ nuclei. The orbitals are
labelled by the approximate asymptotic quantum numbers, 
$\Omega[{\cal N} n_z \Lambda]$, also for $\gamma \neq 0$, even though
 $\Omega$, which is the projection of $j$, is not preserved in
this case. 
The upsloping orbitals emerging from the subshells below the $N=82$,
which are important when building strongly collective bands,
are highlighted. Note how the 1/2[400] and 3/2[402] orbitals repel
each other with increasing axial asymmetry, thus inducing triaxial shape
in configurations with holes in the 3/2[402] orbital. The $N=92$ shell
gap of the TSD band in $^{163}$Lu is marked out and it is then shown how the
TSD bands in $^{161,165,167}$Lu are formed from holes or particles in the
valence orbitals.
}}
\end{figure}

As seen in Fig. 5(a) in Ref.~\cite{Oll09} (and in Fig. \ref{9} below drawn at
a somewhat larger deformation), the proton configuration with two $h_{9/2}$ and one
$i_{13/2}$ proton is favoured for TSD deformations ($\varepsilon_2 \sim 0.37$, $\gamma
\sim 20^{\circ}$) for frequencies up to
$\hbar \omega \sim 0.6$ MeV.
Indeed, according to our calculations, this is the proton configuration, 8(21),
for the lowest calculated TSD bands in the $^{161-167}$Lu isotopes. In order to understand the
neutron configurations, Fig.~\ref{spn} is instructive. 
Starting from the left, it shows the single-neutron
orbitals for prolate shape in the range $\varepsilon_2 = 0.09-0.25$,
then for $\varepsilon_2 = 0.25$ as a
function of axial asymmetry $\gamma$ and finally for constant $\gamma=
20^{\circ}$, again as a function of $\varepsilon_2$. The neutron configurations of the TSD bands
in the Lu isotopes with $N=90-96$ are then illustrated at $\varepsilon_2 \approx 0.40$
(and $\gamma=20^{\circ}$). The gap indicated for $N=92$ is responsible
for the $^{163}$Lu configuration which has two holes in ${\cal N}=4$ and two holes
in $h_{11/2}$ ${\cal N}=5$ orbitals combined with six particles in ${\cal N}=6$ orbitals, i.e. the
configuration (22)6. As discussed e.g. in Ref.~\cite{Pau07}, the holes in the
upsloping ${\cal N}=4$ and $h_{11/2}$ orbitals are very important for the formation
of collective bands, where it is the coupling within the ${\cal N}=4$ orbitals which
induces the triaxial shape according to the
mechanism described in Refs.~\cite{Pau07,Lar74}.

\subsection{Observed and calculated total energies} 
\begin{figure*}[ht]
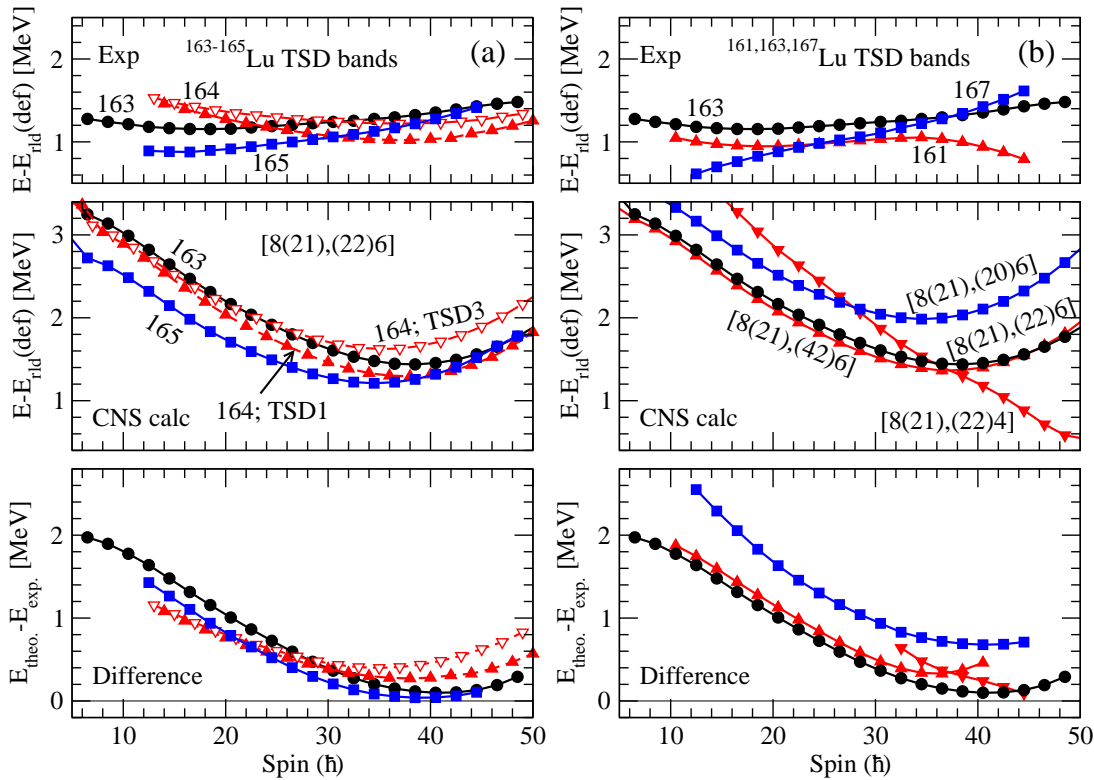

\centering\includegraphics[clip=true,width=0.4\textwidth,angle=0]{expth1-63-65}
\centering\includegraphics[clip=true,width=0.4\textwidth,angle=0]{expth1-61-3-7}
\caption{\label{expth} {\small(Color online) The observed energies of selected TSD bands for Lu
isotopes are shown relative to  the rotating liquid drop energy in the upper
panels, with the calculated bands assigned to them in the middle panels and the
difference between calculations and experiment in the lower panels. 
The TSD bands for $^{163,164,165}$Lu are shown in the panels to the left and
those of $^{161,163,167}$Lu in  the panels to the right, i.e. $^{163}$Lu is
shown in both cases to facilitate the comparison.
Solid lines correspond to positive parity configurations and broken lines correspond to negative parity.
Similarly, solid symbols correspond to signature $\alpha = 1/2$ ($\alpha =
0$ for $A$ even) and open
symbols correspond to signature $\alpha = -1/2$ ($\alpha = 1$). 
Note how
these differences are almost identical for most of the bands where the  differences
between experiment and calculations for the low spin states is understood from
the neglect of pairing correlations in the CNS calculations. The experimental
data 
are taken from Refs.~\cite{16103,16106,16302,16499,16407,GS,16705}
}}
\end{figure*}

Adding one or two neutrons, Fig.~\ref{spn} suggests that the
most favoured configurations for $^{164,165}$Lu will be formed if these neutrons are
placed in the 5/2[523] orbital, where thus two bands with different signature
are formed in $^{164}$Lu. 
In Fig.~\ref{expth}(a), where the observed \cite{GS,16302,16407} 
and calculated bands are compared,
it is
the lowest TSD band in the respective nuclei and in addition band TSD3 in
$^{164}$Lu which are assigned to the configurations discussed above. Note that
contrary to Ref. \cite{16407}, we have assumed that this TSD3 band has
negative parity. The assignment in  Ref. \cite{16407} is based on Ref. 
\cite{16499} where band TSD3 is given positive parity based on the assumption
that it is unlikely with a stretched $M1$ transition with such a high energy
as 1532 keV. We find this conclusion questionable because in the decay of
TSD1, such transitions with 1452  keV and 1541 keV have been
observed in Ref. \cite{16499} and Ref. \cite{16407}, respectively.
Indeed, the similar decays of the TSD1 and TSD3 bands rather suggest to us
that they have the same parity and this conclusion gets additional strong
support from the comparison with calculations, indicating that these two
bands are signature partners.

For $^{163}$Lu, 
the difference between calculations and experiment 
shown in the lower panel of Fig.~\ref{expth}(a) is
close to zero at high spin, where pairing correlations which are not included
in the CNS formalism should be small. The differences are then getting larger
at lower spin values, indicating the increasing importance of the pairing
correlations. The curves for $^{165}$Lu are similar 
leading to close to identical difference curves in the lower
panel of Fig.~\ref{expth}(a). The similarities between the observed bands 
indicate that the orbital
which is occupied in $^{165}$Lu but not $^{163}$Lu is not strongly deformation
polarizing and not giving any large  spin contribution, as is the case for
the 5/2[523] orbital, which is selected in the calculations. The two
bands in $^{164}$Lu, come close to the average of the $^{163}$Lu and $^{165}$Lu bands
at high spin in  Fig.~\ref{expth}(a). Indeed, this is the case for all observed spin values
in the (unpaired) calculations, while at lower spin values the odd-N energies
come higher in experiment. This is what would be expected from a smaller
pairing energy in the odd compared with the even neutron systems and it
should  even be possible to get an idea of the strength of the pairing
correlations from this comparison. Furthermore, the calculations predict the
correct signature for the favoured bands in $^{164}$Lu. This gives additional
support to the present assignments even though the splitting is somewhat
overestimated in the calculations.

Fig.~\ref{spn}
suggests that the additional holes in $^{161}$Lu relative to $^{163}$Lu should
be placed either in the $i_{13/2}$, 5/2[642] orbital or in the 1/2[400]
orbital. The result of the detailed calculations, see Fig.~\ref{expth}(b), is
that the latter deexcitation, i.e. the neutron configuration (42)6 is favoured
for lower spin values while the former deexcitation, i.e. the neutron
configuration (22)4 is favoured for higher spin values. Indeed, it appears that
this agrees with experiment \cite{16103,16106} because in the observed band, one can see a smooth
crossing for spin values $I=30-40$, where the two unpaired configurations
cross. Thus, with this assignment and with our choice of spin values
for the $^{163}$Lu band, the difference curve in the lower panel of
Fig.~\ref{expth}(b) have almost the same shape as for  
$^{161}$Lu (where we have chosen an excitation energy of the
unlinked band in $^{161}$Lu similar to that for the 
$^{163}$Lu band). 
Furthermore, with pairing included, the crossing between the
neutron (42)6 and (22)4 configurations will be seen as a smooth paired
crossing within the $i_{13/2}$ orbitals~\cite{Ma}.
Note that the two neutrons which are shifted
from down-sloping to up-sloping orbitals lead to a considerably larger 
deformation for the (42)6 configuration, $\varepsilon_2 \sim 0.43$, $\gamma
\sim 23^{\circ}$, than for the (22)4 configuration, $\varepsilon_2 \sim 0.37$, 
$\gamma \sim 20^{\circ}$. This latter deformation is typical for the yrast
TSD bands in the other Lu isotopes with $N= 92-96$. 

Coming to $^{167}$Lu, Fig.~\ref{spn} indicates that the two additional neutrons compared
with $^{165}$Lu might be put in the 11/2[505] orbital or in the 7/2[633]
orbital. However, the detailed calculations show that the latter configuration
is much less favoured for spin values above $I=30$ in accordance
with the general experience that it becomes energetically expensive to build
spin in configurations of high-$j$ shells which are half-filled or more than 
half-filled, see e.g. Fig. 12.11 of Ref.~\cite{book}. 
As seen in Fig.~\ref{expth}(b), the energy vs. spin dependence of the 
(20)6 configuration in $^{167}$Lu is close to that of the (22)6 configuration in
$^{165}$Lu while the calculated energy is considerably higher in $^{167}$Lu than in
$^{165}$Lu in disagreement with experiment. This discrepancy would disappear if
the $h_{11/2}$ subshell was lowered by a few hundred keV.

There are a few more observed TSD bands in Lu nuclei which we have not
considered here. Thus, there are three unlinked bands in $^{162}$Lu~\cite{16103}. 
It appears to be easy to assign spins and excitation energies 
to these bands so
that they agree with calculations, but these assignments would be very 
tentative. One could note however that the beginning of a band-crossing
is observed in the TSD3 band which appears to be very similar to the 
band-crossing in $^{161}$Lu suggesting a similar origin and thus an
appreciable deformation change also in $^{162}$Lu. Another band which
we have not discussed here is TSD2 in $^{164}$Lu~\cite{16407}. One could expect a
neutron configuration with all orbitals up to the $N=94$ gap in occupied
but with a hole in the unfavoured 5/2[642] orbital, see Fig. \ref{spn}. 
Indeed, the parity and
signature of the observed band agrees with this assignment but the 
curvature of the $E$ vs. $I$ function of the calculated configuration appears too
large. In addition, the observed band appears to go through a smooth
band-crossing which is not easy to explain. There is an interesting branch
of band TSD1 at high spin which has a larger alignment and is referred
to as X2~\cite{16407}. This branch might be assigned to the configuration with the
valence neutron excited from the favoured 5/2[523] orbital to the
favoured 1/2[770] $i_{13/2}$ orbital, see  Fig.~\ref{spn2} below.
In addition, there are several bands assigned as wobbling excitations
in the odd Lu isotopes
which will of course not be described by any CNS configuration.

Fig.~\ref{spn} is drawn at no rotation, $\omega = 0$, and is thus mainly
helpful for the
understanding of configurations at low or intermediate angular momentums.
In order to get an understanding of the configurations which are favoured
at higher angular momentums, it is more instructive to draw a single-particle diagram at
$\omega > 0$, which will lead to a more complicated diagram because the orbitals
will split into two branches with signature $\alpha = 1/2$ and $\alpha =
-1/2$. Such a diagram is provided in Fig.~\ref{spn2}. It suggests
\begin{figure}[ht]
\centering\includegraphics[clip=true,width=0.47\textwidth,angle=0]{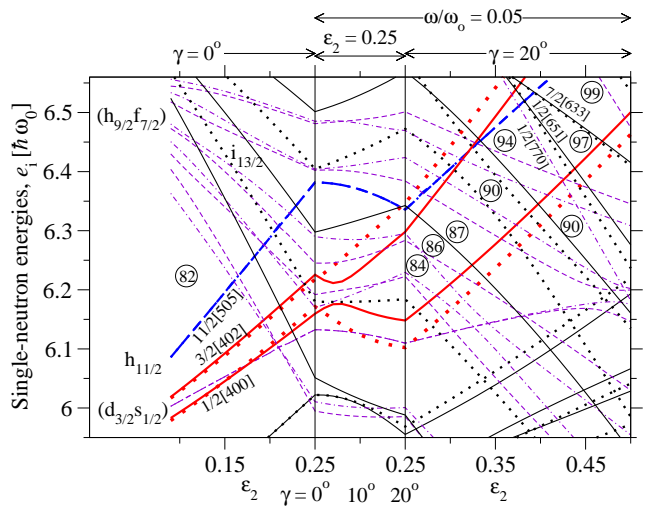}
\caption{\label{spn2} {\small(Color online) Similar to Fig.~\ref{spn} but with rotation added, i.e.
a rotational frequency which increases linearly up to $\omega/\omega_0 = 0.05$
in the `Nilsson diagram' to the left and then keeping this value of the
rotational frequency when axial asymmetry is added. Solid and dotted lines are
used for positive parity and dashed and dot-dashed lines for negative parity,
where dots are used for signature, $\alpha = -1/2$.
}}
\end{figure}
that the favoured configurations for $N=90-94$, i.e. for $^{161-165}$Lu, will be
about the same as for $\omega = 0$, but for $N=96$ ($^{167}$Lu), it will be more favourable
to put the two extra neutrons in the lowest 1/2[770] orbital or in the 1/2[651] 
orbital (of $j_{15/2}$ and $i_{11/2},g_{9/2}$ origin, respectively). This is also in
agreement with the detailed calculations which shows that such a configuration
becomes favoured in energy at triaxial shape above $I\sim35$ when
combined with the same favoured proton configuration as for the lower spin
states 8(21). At these higher frequencies and deformations, it will however
be favourable if also the deformation driving second proton $i_{13/2}$ orbital will be
occupied leading to the favoured 8(22) configuration for $^{168}$Hf which will
be discussed below.

\begin{figure}
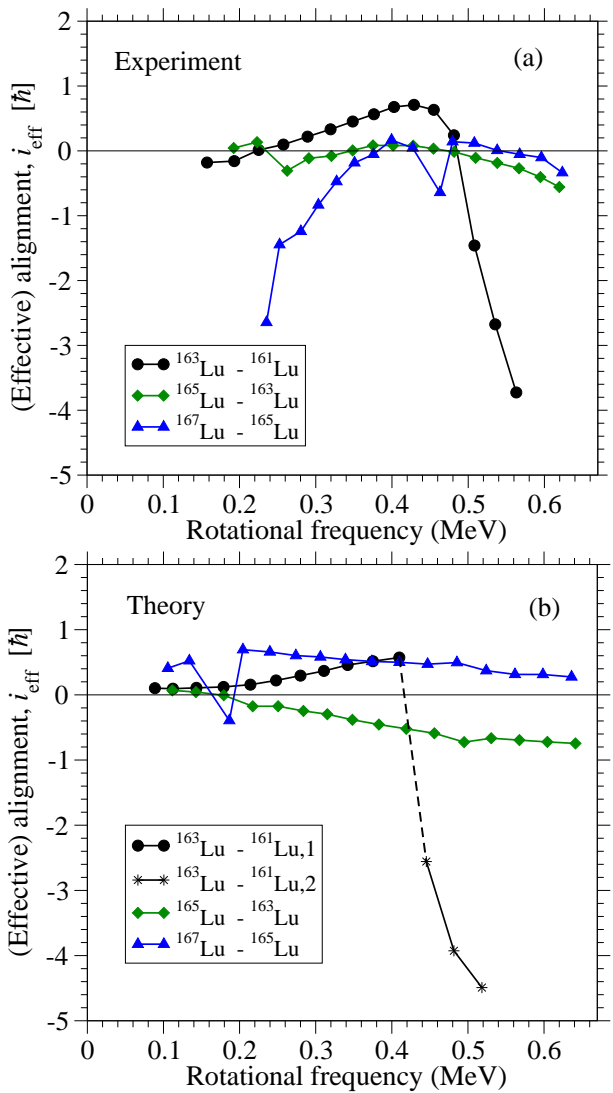

\begin{centering}
\includegraphics[scale=0.40,clip=true]{ieff-lu-exp}
\includegraphics[scale=0.40,clip=true]{ieff-lu-theo}
\caption{\label{ieff}{\small(Color online) Effective alignment, $i_{eff}$,
    for Lu isotopes, extracted from (a) experiment and (b) corresponding calculated bands. }}
\end{centering}
\end{figure}

\subsection{Effective alignments, $\i_{eff}$}
In our analysis of TSD bands in Lu isotopes, 
we will also consider the differences of spin, $I$, at a constant frequency, $\hbar\omega$, and compare the experimental and theoretical data.
This quantity referred to  as the effective alignment, $i_{eff}$ has been
very important for the classification of the SD bands in the $A=150$ region,
see e.g. \cite{Rag93,Haa93,Afa98}. It is a direct measure of the
contribution from different Nilsson
orbitals. It is mainly useful when pairing can be neglected but for the Lu
bands, the pairing correlations are rather small and we can furthermore assume 
that pairing gives about the same contribution if the comparison is limited
to the odd isotopes with an even number of neutrons.
%
Thus, effective alignments of neutron orbitals for the lowest TSD bands in Lu
nuclei are shown as a function of rotational frequency $(\omega=E_{\gamma}/2)$, 
for the experimental bands
in Fig.~\ref{ieff}(a) and for the theoretical 
configurations assigned to these bands in Fig.~\ref{ieff}(b).
Note that in this case, $i_{eff}$ is a measure of the spin contribution
from a pair of particles in the respective orbitals.
  
The general agreement between experiment and theory in Fig.~\ref{ieff}   
indicates that we do understand which orbitals are filled in the lowest TSD
bands in the odd Lu isotopes. 
The spin contribution of the orbital which is being occupied when 
going from $^{161}$Lu to the $^{163}$Lu is very small and positive 
at $\hbar\omega\lesssim0.5$ MeV 
but it changes for $\hbar\omega\gtrsim0.5$ MeV where $i_{eff}$ turns
negative. The calculated $i_{eff}$ shows the same
feature which can be traced back to a change of structure in $^{161}$Lu 
from [8(21),(42)6] to [8(21),(22)4] at $\hbar\omega\sim0.5$ MeV.
The value of $i_{eff}$ when comparing the bands in $^{163}$Lu and $^{165}$Lu
is close to zero but rather negative, corresponding to a small 
negative  spin contribution from the
orbital which becomes occupied. This orbital is located in the middle
of the $h_{9/2}f_{7/2}$ subshells and is labelled 5/2[523] in Fig. \ref{spn}. 
%
When two neutrons are added to $^{165}$Lu, a spin contribution close to zero
is obtained in both experiment and calculations for $\hbar \omega \gtrsim 0.5$ MeV.
This agreement supports the assignment
that it is the highest $h_{11/2}$ orbital, 11/2[505], which is being occupied.
Note that this upsloping orbital will have a strong shape polarization,
i.e. the shape change will have an important contribution to $i_{eff}$,
see e.g. \cite{Rag90}. The
fact that calculations and experiment diverge at smaller frequencies could
be caused by increasing pairing correlations so that the  assumption that
an orbital is either filled or empty is strongly violated.  


The present calculations show that the standard CNS formalism provides a 
reasonable interpretation for the TSD bands in Lu isotopes. However, it is 
questionable whether this approach, including approximations pointed out 
in sect.~\ref{model},  is suitable to study also the 
TSD bands in $^{168}$Hf which have a larger deformation. In the next section,
these approximations will be tested on $^{168}$Hf.

\section{Analysis of specific features of the CNS formalism}
\label{tests}

\begin{figure*}
\begin{centering}
\includegraphics[scale=0.39]{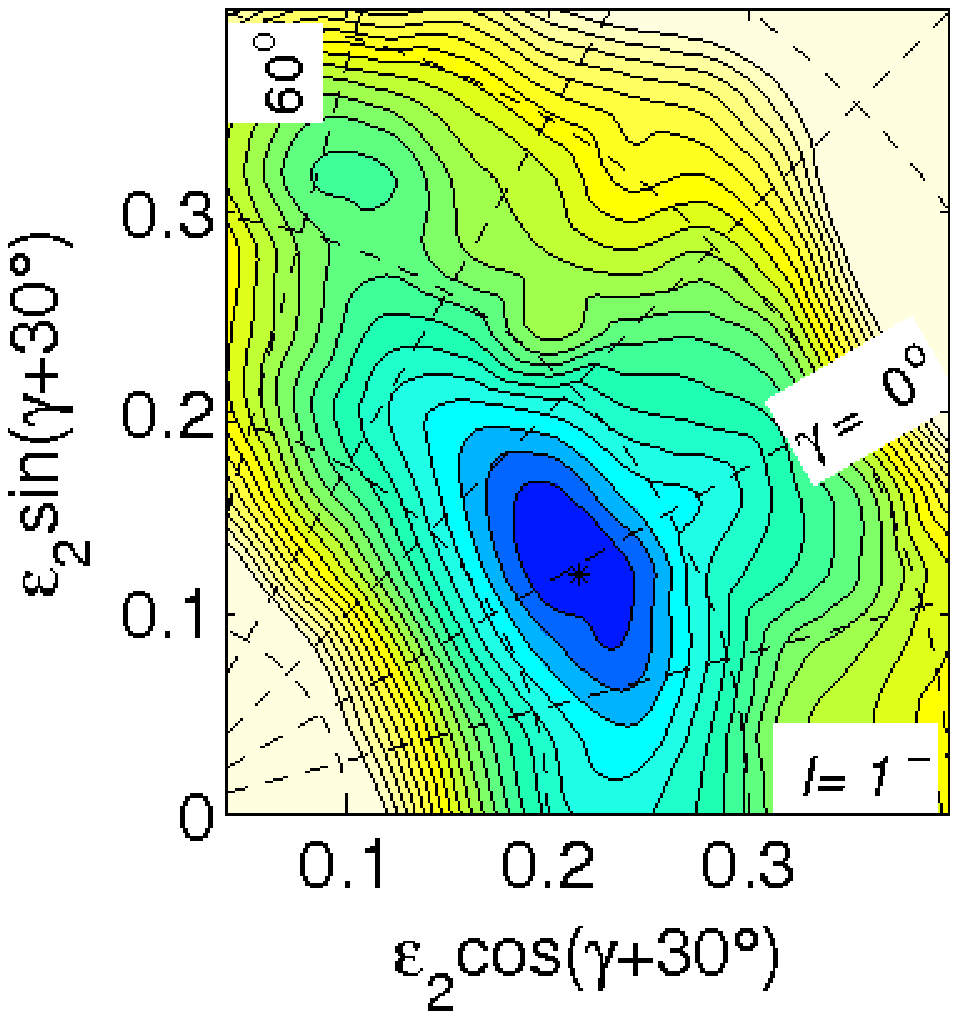}\includegraphics[scale=0.39]{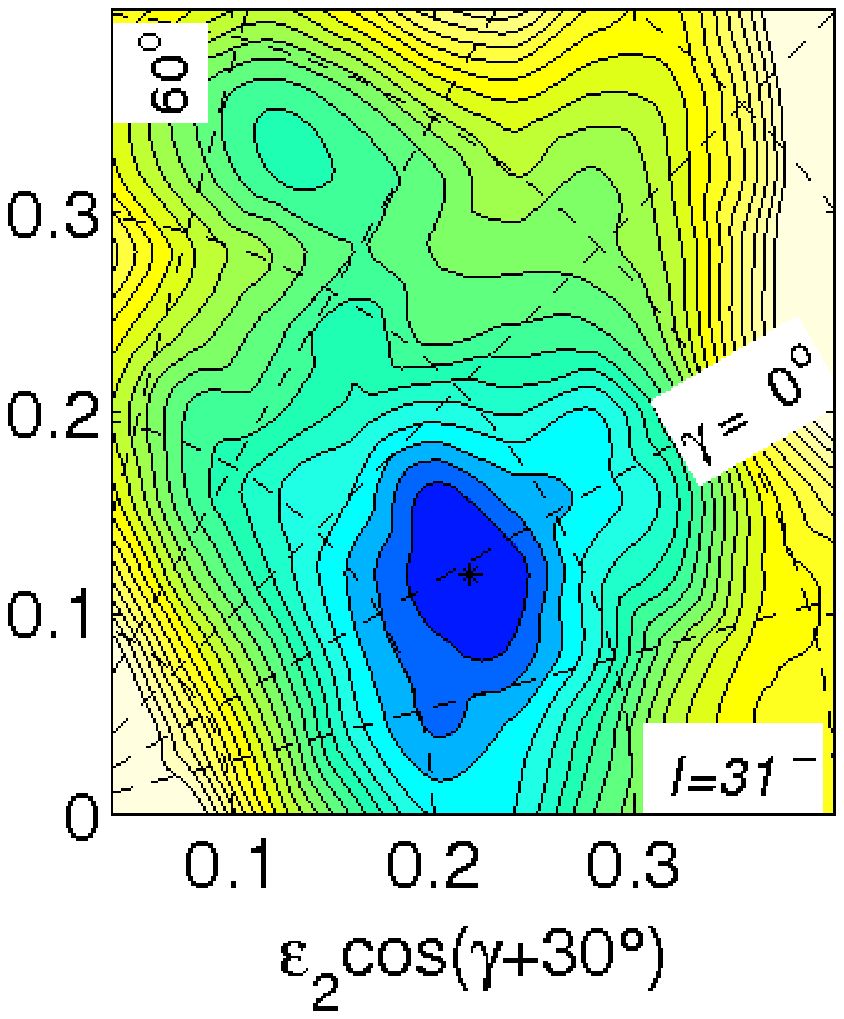}\includegraphics[scale=0.39]{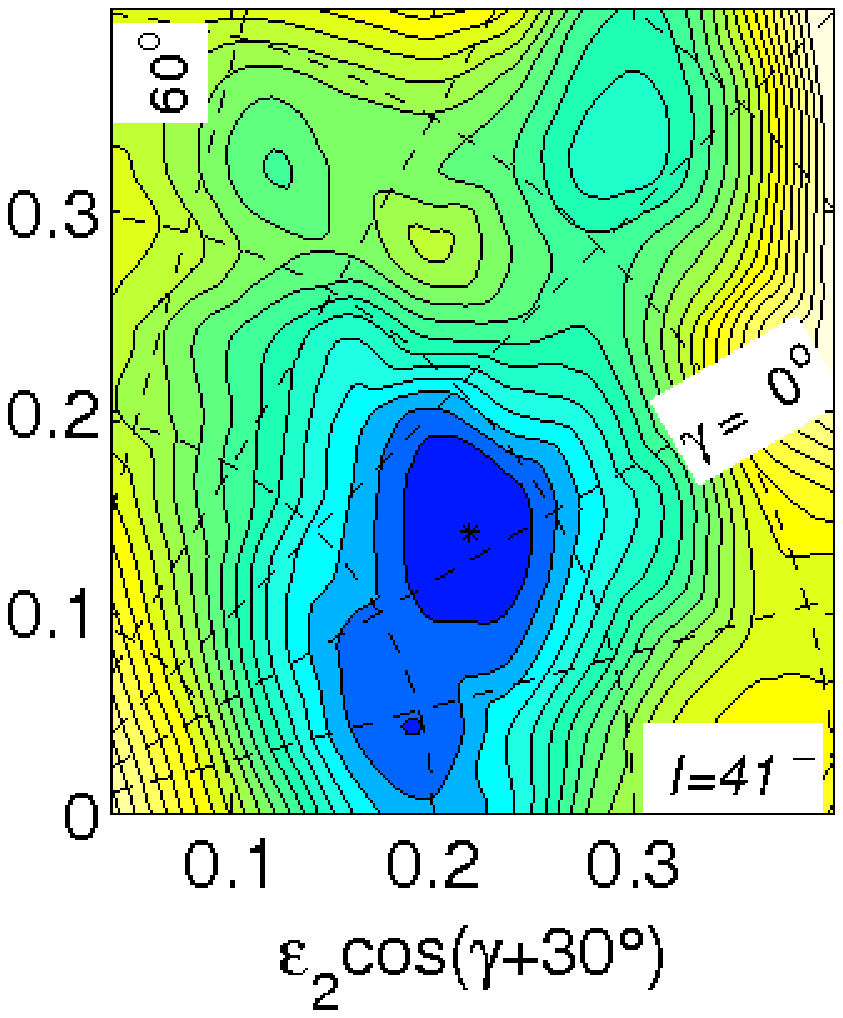}\includegraphics[scale=0.39]{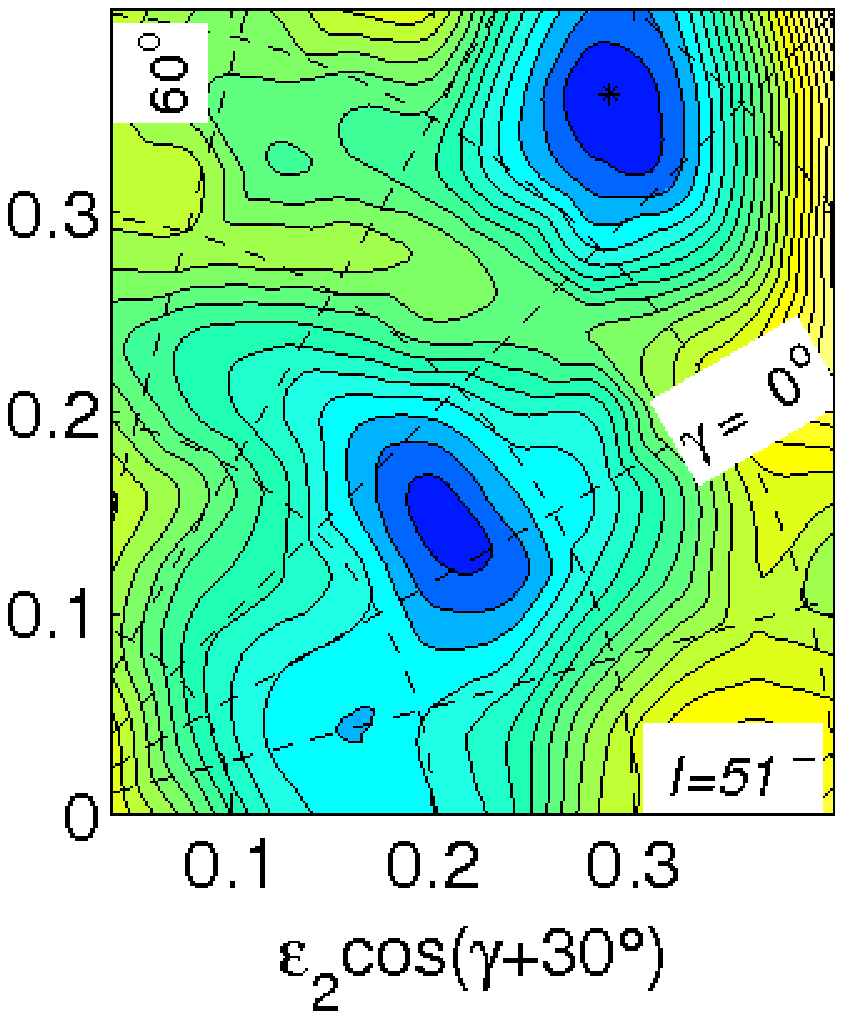}\includegraphics[scale=0.39]{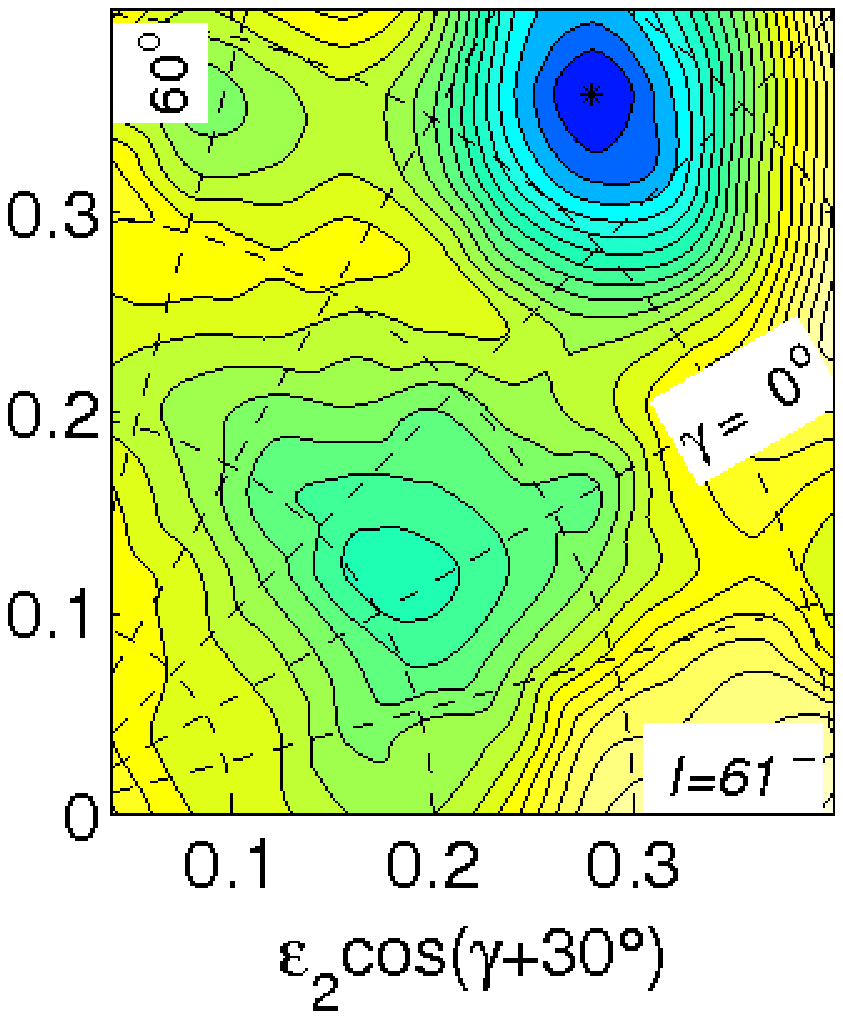}
\caption{\label{posu}{\small(Color online) Calculated potential-energy surfaces versus quadrupole deformation
$\varepsilon_{2}$ and the triaxiality parameter 
$\gamma$ of $^{168}$Hf
with $(\pi,\alpha)=(-,1)$ for spins $I=1,31,41,51,61$. Contour
lines are separated by 0.25 MeV and the $\gamma$ plane is marked at $15^{\circ}$
intervals. Dark regions represent low energy with absolute minima
labeled with a dot.}}
\end{centering}
\end{figure*}

Representative potential energy surfaces (PES) 
with $(\pi,\alpha)=(-,1)$ for spins $I=1,31,41,51,61$
are displayed in Fig.~\ref{posu} for $^{168}$Hf. Similar behavior is also
found for the other $(\pi,\alpha)$ combinations. At low spins, from
$I=1$ to $I=31$, the lowest energy minimum in the PES's corresponds
to a almost prolate shape at $(\varepsilon_2,\gamma)\sim(0.23,0^{\circ})$.
As the angular momentum increases, this minimum migrates to a somewhat
larger deformation; for example  $(\varepsilon_2,\gamma)\sim(0.26,\,3^{\circ})$ at spin 
$I=41$.
For spin values $I\gtrsim50$, the minimum energy corresponds to a TSD shape at
the deformation $(\varepsilon_2,\gamma)\sim(0.44,\,20^{\circ})$. 

\subsection{Off-shell matrix elements and More shells}
\label{off}
As it has been pointed out in sect. \ref{model}, all off-shell 
elements in the rotating basis $| n_{x}n_{2}n_{3}\Sigma \rangle$ are small
and  it is therefore natural to neglect them.
If the  off-shell matrix elements are included, the shell number ${\cal N}_{rot}$
will not the good quantum number and the rotating basis functions lose 
their advantage to diagonalize the Hamiltonian matrix. 
It is then easier to use the stretched
spherical harmonic basis functions $| N_t \ell_t j_t \Omega_t \rangle$ 
which are eigenkets of the spherical harmonic oscillator Hamiltonian 
$h_{HO}(\varepsilon_2,\gamma)$, the square of the stretched angular 
momentum $j_t^2$ and its projection, $j_{z,t}$. 
With these basis functions, the 
cranking term couples between basis state of the shells
${\cal N}_{t}$ and ${\cal N}_{t}\pm2$ which have the same signature. 

\begin{figure}[ht]
\begin{centering}
\includegraphics[clip=true,scale=0.32]{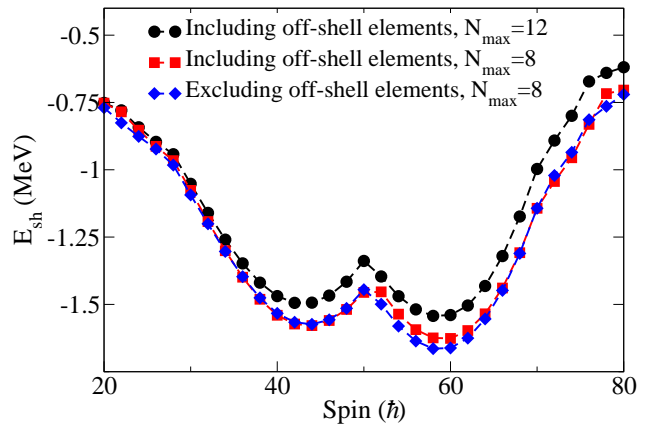}
\caption{\label{3}{\small(Color online) The shell energy for the configuration {[}8(22),(22)6(11){]} with 
deformation parameters $\varepsilon_{2}\sim0.43,\gamma\sim20^{\circ},\varepsilon_{4}\sim0.028$ in the $^{168}$Hf nucleus.
The circle symbol shows the calculations including the off-shell
elements and ${\cal N}_{max}=12$, the square symbol including the off-shell elements
and ${\cal N}_{max}=8$ and the diamond symbol excluding the off-shell elements and
${\cal N}_{max}=8$. }}
\end{centering}
\end{figure}

When calculating the total energy, we need the shell energy 
and the rotating liquid drop energy 
(Eq.~(\ref{eq:sh})). The addition of the off-shell 
elements will only effect the shell energy. 
As illustrated in Fig.~\ref{3}, 
the shell energies obtained from the diagonalization of the 
Hamiltonian in the two cases come very close for all spin values
at a large triaxial
deformation with a typical (see below)
hexadecapole deformation, $\varepsilon_4 = 0.028$. 
Note that even though the coupling between the ${\cal N}_{rot}$ shells is 
neglected in the rotating basis, the $j_{x}$ term is still
fully accounted for because it is included in the basis. This
is contrary to the stretched basis where the finite basis size
corresponds to a (small) approximation. With more shells included, this
approximation will be negligible.

In the standard CNS calculations, all shells having the principal quantum
number less than or equal to ${\cal N}_{max}=8$
are included in the diagonalization.
The important question is now if more shells are needed in order to reproduce
the solution accurately enough for a heavy nucleus like $^{168}$Hf.
Naturally, the required value of ${\cal N}_{max}$ depends on particle number, the
shape of the potential to be diagonalized and for the stretched basis also on
the rotational frequency, $\omega$.  To illustrate the importance of the
cut-off error, the yrast energy was calculated including off-shell couplings
with ${\cal N}_{max}=12$, i.e. with four added shells.  For the specific deformation
illustrated in Fig.~\ref{3}, it turns out that the energy of the yrast line
with ${\cal N}_{max}=12$ does not decrease relative to the calculation with
${\cal N}_{max}=8$ but it rather increases. The reason is that with the increase of
the number of shells, both the total discrete and smoothed energy
decrease. The total discrete single-particle energy with ${\cal N}_{max}=12$ differs from that
 with ${\cal N}_{max}=8$ by about 30 keV for spins $I \lesssim 40$ and 120 keV for
spins $I\gtrsim40$. Since the corresponding smoothed single-particle energy is
shifted by about 90 keV at spins $I\lesssim40$ and 260 keV at spins
$I\gtrsim40$, the resulting shell energy, 
\begin{equation}
E_{sh}=\sum_{i}e_{i}- \langle \sum_{i}e_{i} \rangle        
\label{eq:4}
\end{equation}
differs only by $\sim 40$ keV at spins $I\lesssim40$ and $\sim 140$ keV $I\gtrsim40$ 
from the corresponding value with ${\cal N}_{max}=8$, see Fig.~\ref{3}.
Thus for the equilibrium deformations
of $^{168}$Hf in an extended spin range,
the cut off at ${\cal N}_{max}=8$ introduces only small changes in $E_{sh}$ which
are essentially negligible compared with other uncertainties.
\subsection{Minimization in five dimensions}
\label{min}
\begin{figure*}
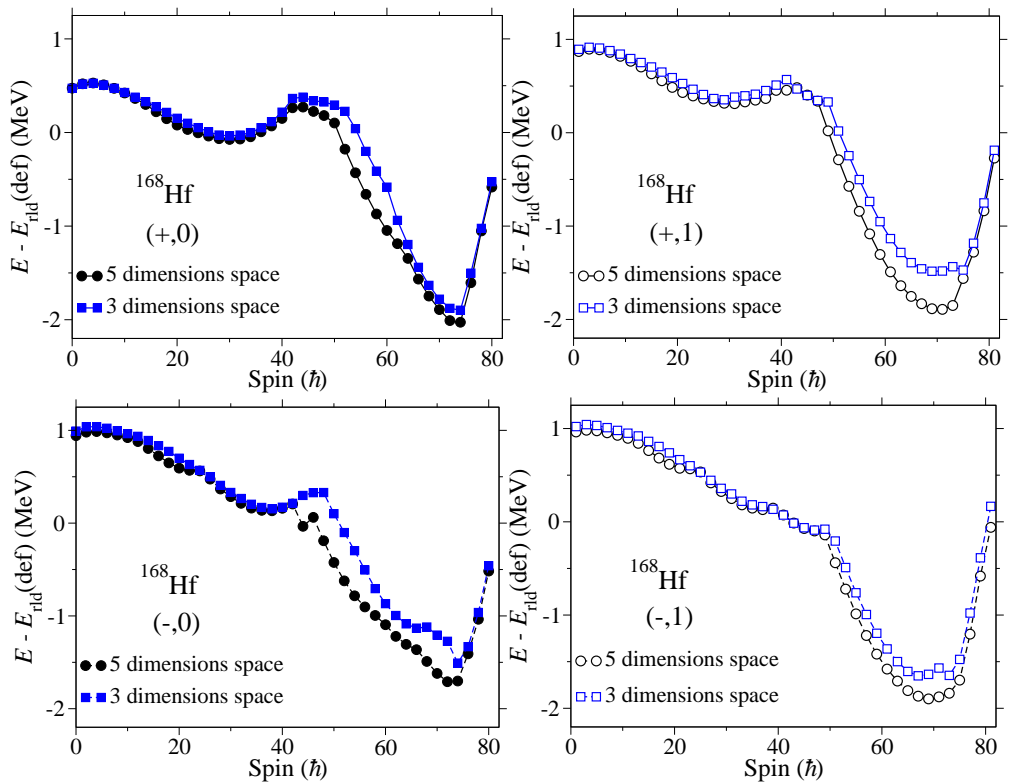

\begin{centering}
\includegraphics[clip=true, scale=0.24]{Hf168-scpp}\includegraphics[clip=true, scale=0.24]{Hf168-scpm}
\includegraphics[clip=true, scale=0.24]{Hf168-scmp}\includegraphics[clip=true, scale=0.24]{Hf168-scmm}
\caption{\label{4}{\small(Color online) The $^{168}$Hf yrast energies relative to a rotating liquid drop
energy E$_{rld}$ as a function of spin $I$ for the four combinations
of parity and signature, $(+,0),(+,1),(-,0)$ and $(-,1)$. The circle symbol
 shows the minimum energy in $(\varepsilon_2,\gamma,\varepsilon_{40},\varepsilon_{42},\varepsilon_{44})$
space of deformation and the square symbol in $(\varepsilon_2,\gamma,\varepsilon_{4})$. Solid lines correspond to positive
parity configurations and broken lines correspond to negative parity.
Similarly, solid symbols correspond to signature $\alpha = 0$ and open
symbols correspond to signature $\alpha = 1$. The steep increase at spin values
$I > 72$, which are most apparent for the 
$(+,0)$ and $(-,0)$ yrast energies is mainly caused by the rotating
liquid drop reference energy which shows a discontinuity when the
equilibrium shape moves away from the $\gamma = 60^{\circ}$ axis,
the superbackbend according to Ref.~\cite{GA} (see sect. \ref{ref}).}}

\end{centering}
\end{figure*}
In general, for axial symmetric shapes it is only the $\varepsilon_{40}$ 
(with quantization around the symmetry axis) shape
degree of freedom which is expected to be of major importance because the
energy is even (independent of the sign)
in $\varepsilon_{42}$ and $\varepsilon_{44}$. This is only
valid at no rotation around the perpendicular axis but if the rotational
frequency is not extremely high, it is still expected that only the
$\varepsilon_{40}$ degree of freedom will be of major importance.
Furthermore, shapes corresponding to small quadrupole deformations, 
are never far away from a symmetry axis in the 
$(\varepsilon_{2}, \gamma)$-plane so it should be sufficient to
minimize the energy in only one $\varepsilon_{4}$ degree of freedom
also in this case. This is supported by studies of the smooth terminating
bands in $^{109}$Sb~\cite{Sch96, AA} where the energy is lowered by less than $\sim 50$ keV
when it is minimized in three $\varepsilon_{4}$ degrees of freedom
~\cite{Sil10}. For a triaxial
shape and large quadrupole deformation on the other hand, the full
minimization in the $\varepsilon_{4i}$-parameter space might be more important.


In order to make a full minimization in the five dimensional deformation space,  
the total energy of $^{168}$Hf is calculated at the following grid points:
\begin{eqnarray*}
x & = & 0.18[0.02]0.44 \\
y & = & 0.08[0.02]0.42 \\
\varepsilon_{40} & = & 0.005[0.01]0.045 \\
\varepsilon_{42} & = & -0.02[0.01]0.02 \\
\varepsilon_{44} & = & -0.01[0.01]0.03, 
\end{eqnarray*}
where $(x,y)$ are 
Cartesian coordinates in the $(\varepsilon_2, \gamma)$-plane. The
$(x,y)$-coordinates are connected with $(\varepsilon_2, \gamma)$ 
by the expressions
\begin{eqnarray}
 x=\varepsilon_2\cos(\gamma+30^{\circ}), \;\;\;  y=\varepsilon_2\sin(\gamma+30^{\circ})\nonumber
\end{eqnarray}
In our numerical calculations, the quantization axis coincides with the
rotation axis to simplify the diagonalization. Therefore, $\gamma$ should
be replaced by $(\gamma+120^{\circ})$ in Eq.~(\ref{eq:3}),
when defining the $\varepsilon_{4i}$ parameters.
With this definition, we relabel the principal axis 
but the same nuclear shapes are formed in the $\varepsilon_{4i}$-space.
Especially, it is for rotation around the symmetry axis ($(\gamma =
60,-120^{\circ}$) that axially symmetric shapes are formed with only
$\varepsilon_{40} \neq 0$ while axially symmetric shapes at 
$\gamma = 0^{\circ}$ are described by all $\varepsilon_{4i} \neq 0$. 
\begin{figure*}
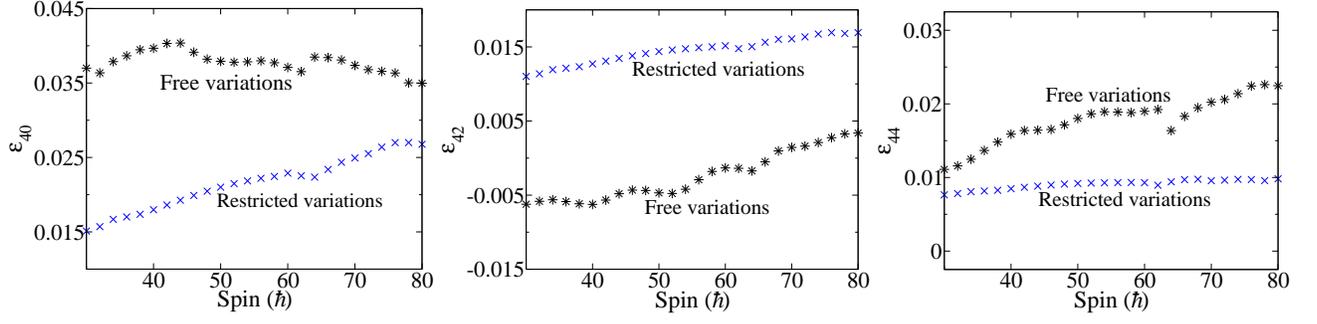

\begin{centering}
\includegraphics[clip=true, scale=0.18]{eps4022611}
\includegraphics[clip=true, scale=0.18]{eps4222611}
\includegraphics[clip=true, scale=0.18]{eps4422611}

\caption{\label{5}{\small(Color online) The $\varepsilon_{4i}$ parameters 
as a function of spin $I$ for the
TSD configuration {[}8(22),(22)6(11){]}. In the calculations, the
triaxiality parameter of Eq.~(\ref{eq:3}) is ($\gamma+120^{\circ}$). The $\times$ symbols are for the minimization
process in the space $(\varepsilon_2,\gamma,\varepsilon_{4})$ while
the $*$ symbols are used for the minimization process in the space
$(\varepsilon_2,\gamma,\varepsilon_{40},\varepsilon_{42},\varepsilon_{44})$. 
}}
\end{centering}
\end{figure*}

 In Fig.~\ref{4},
the $^{168}$Hf yrast energies are drawn relative to a rotating liquid
drop energy E$_{rld}$ as a function of spin $I$ for the four combinations
of parity and signature, 
$(\pi,\alpha) = (+,0),(+,1),(-,0)$ and $(-,1)$. They
are compared with the corresponding energies from the  
minimization in the $(\varepsilon_2,\gamma,\varepsilon_{4})$
parameter space.
In our calculations, the reference energy E$_{rld}$
is minimized in a deformation space $(\varepsilon_{2},\gamma,\varepsilon_{4})$ for 
each spin value. 

 As one can see, at spins $10\lesssim I\lesssim45$, the yrast states
in the deformation space $(\varepsilon_2,\gamma,\varepsilon_{40},\varepsilon_{42},\varepsilon_{44})$
are only a few keV lower in energy than that of in the 
$(\varepsilon_2,\gamma,\varepsilon_{4})$ deformation space.
On the other hand, the gain in energy in the high spin region, $I\gtrsim45$,
is important and amounts to 0.5 MeV at some spin values. 
These findings are consistent with the general expectations discussed
above. Thus, according to the
potential-energy surfaces in the CNS calculations for $^{168}$Hf (see Fig.~\ref{posu}),
the yrast states are built from configurations
which have prolate shape with $\varepsilon_{2}\sim(0.23-0.26)$ for spin values
below $I\sim45$ but at non-axial shape with 
$(\varepsilon_{2},\gamma)\sim(0.44,20^{\circ})$ (TSD shapes)
for spins $I \gtrsim 45$. 
Therefore in the  following, we do the minimization process in the deformation space $(\varepsilon_2,\gamma,\varepsilon_{4})$
to study the bands close to axial shape and in the deformation space $(\varepsilon_2,\gamma,\varepsilon_{40},\varepsilon_{42},\varepsilon_{44})$
to study the TSD bands in $^{168}$Hf.

In order to illustrate the variation of the $\varepsilon_{4i}$
parameters in the two cases, they are drawn in Fig.~\ref{5}
as functions of spin $I$ for the TSD configuration, {[}8(22),(22)6(11){]}.
In the complete minimization, the $\varepsilon_{4i}$
parameters get different values relative to Eq.~(\ref{eq:3}) 
in the full spin range, $I=30-80$.
The value of the $\varepsilon_{40}$ parameter becomes considerably larger,
$\varepsilon_{40} \sim 0.035$ compared with $\varepsilon_{40} \sim 0.020$
in the restricted variation.
The $\varepsilon_{42}$ parameter changes sign over most of the spin range
while the $\varepsilon_{44}$ parameter varies faster
and gets larger values. 

\begin{figure}[ht]
\begin{centering}
\includegraphics[clip=true, scale=0.22]{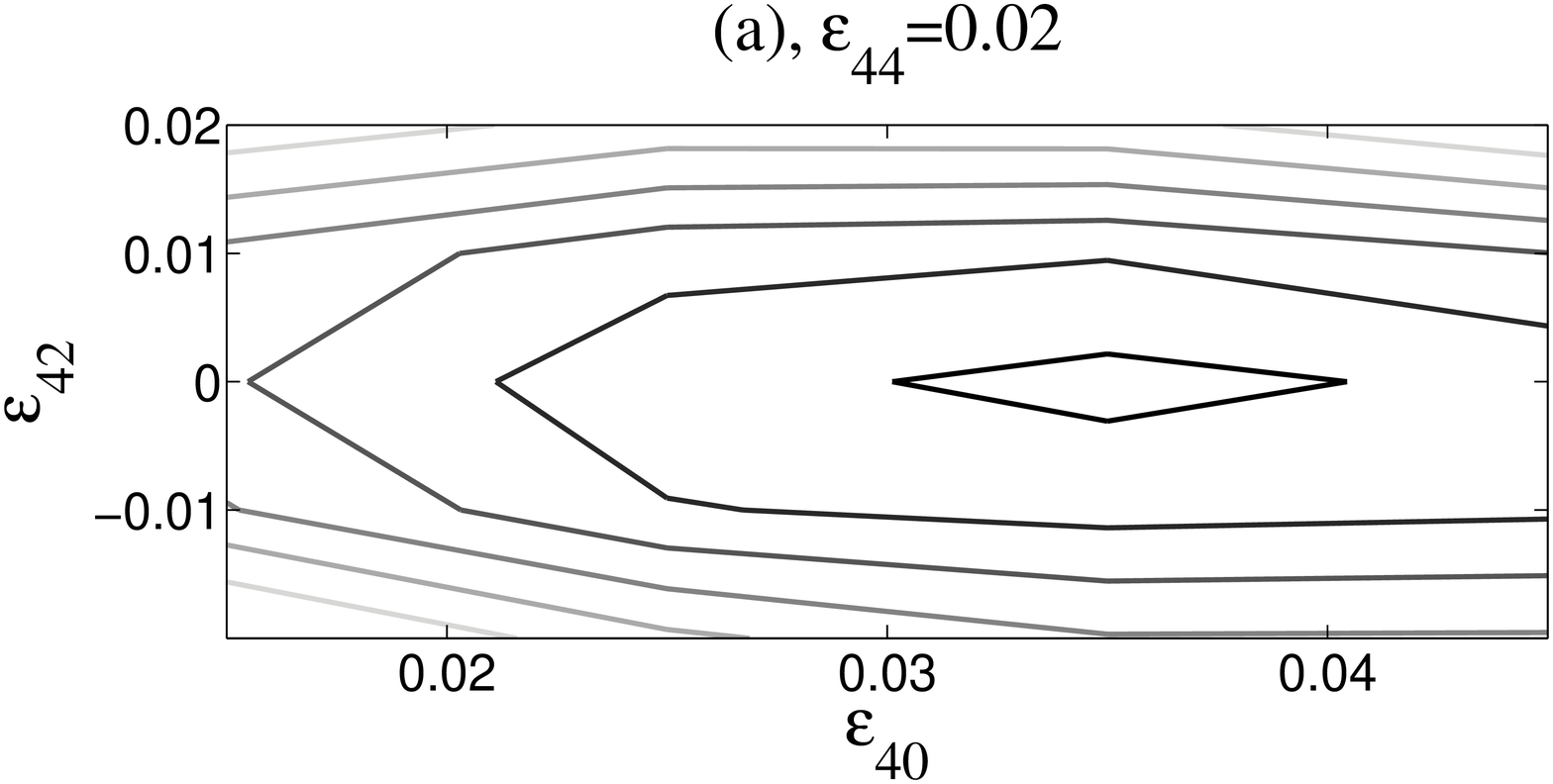}
\includegraphics[clip=true, scale=0.22]{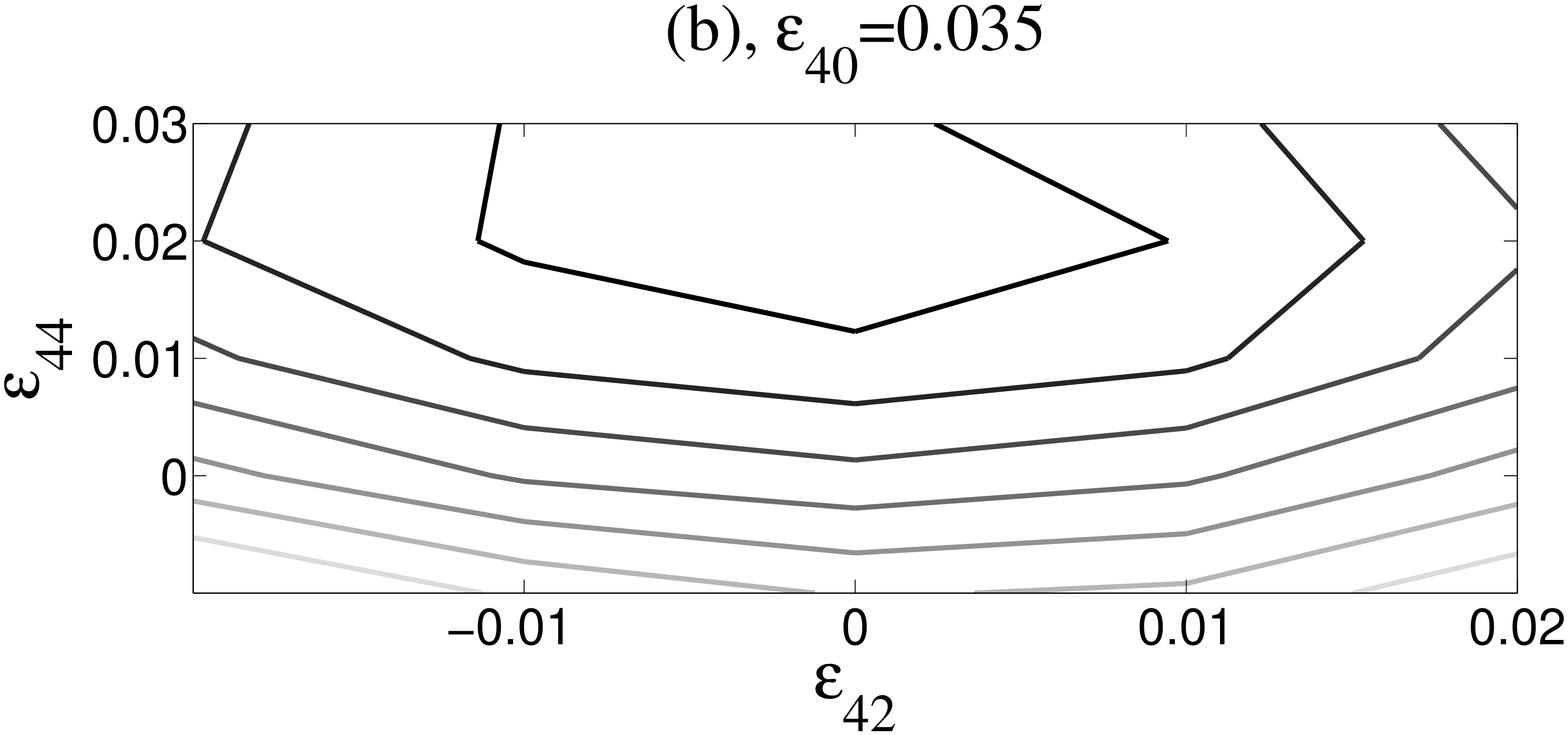}
\includegraphics[clip=true, scale=0.22]{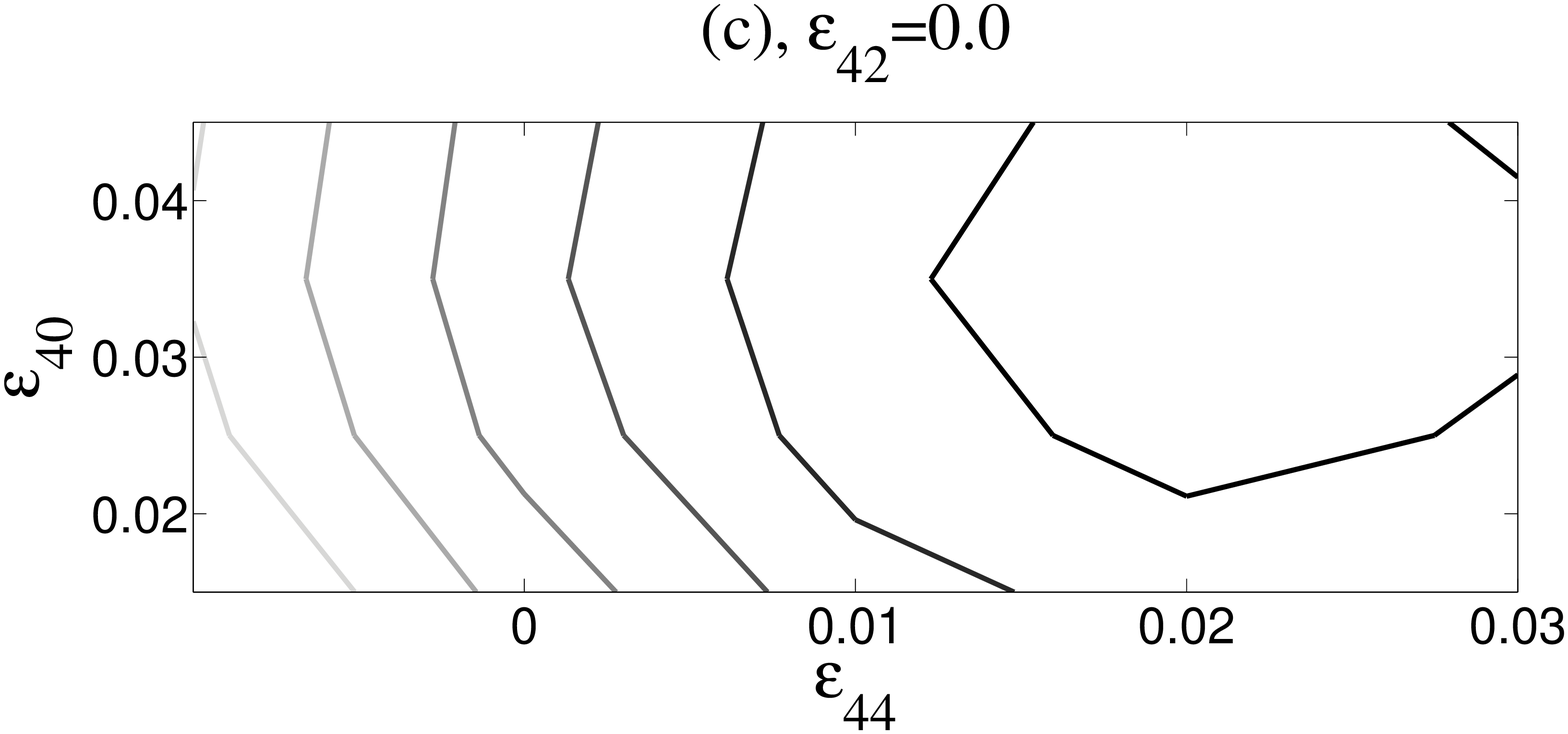}

\caption{\label{6}{\small Energy surfaces shown as functions of two of the
three $\varepsilon_{4i}$ parameters
for the configuration {[}8(22),(22)6(11){]} at spin
$I=50$ and quadrupole deformation parameters $(\varepsilon_{2},\gamma) = (0.43,20^{\circ})$.
The contour line separation is $0.1$ MeV in (a) and $0.2$ MeV in (b)
and (c).}}
\end{centering}
\end{figure}

The discontinuity in the variations of the $\varepsilon_{4i}$ parameters
at spin $I\sim65$ is understood from a crossing of high-$j$ and low-$j$
orbitals in this configuration which is explained below. The energy
surfaces at spin $I=50$ and for the 
same {[}8(22),(22)6(11){]} configuration are shown in Fig.~\ref{6}(a-c),
in the planes $(\varepsilon_{40},\varepsilon_{42})$,
$(\varepsilon_{42},\varepsilon_{44})$ and
$(\varepsilon_{44},\varepsilon_{40})$
for a constant value close to the minimum of the third parameter.
These figures indicate that the total energy is
well-behaved with only one minimum in the
$(\varepsilon_{40},\varepsilon_{42},\varepsilon_{44})$ space.

\subsection{The reference energy}
\label{ref}
In order to highlight the details of high-spin bands, their energy is
often shown relative to a reference. For a long time, the standard choice
of such a reference has been 
$E_{ref} =C \cdot I(I+1)$ MeV/$\hbar^2$, 
where $C$ is a constant \cite{TB}
for a specific nucleus. 
In calculations based on the CNS approach, the constant has generally
been chosen as $C=32.32 A^{5/3}$ MeV \cite{AA}, which means that the reference
energy corresponds to rigid rotation at a prolate deformation,
$\varepsilon = 0.23$, assuming a sharp nuclear radius $r_0 A^{1/3}$
with $r_0 = 1.2$ fm. With this choice, the increase or 
decrease of $E(I) - E_{ref}$
is relevant and it becomes instructive to compare rotational bands in
different mass regions. On the other hand, 
different constants have been used in the literature so one should be careful
before drawing any conclusions from the slope of $E(I) - E_{ref}$ curves.  
For examples, while the $A$-dependent expression 
specified above gives $C=0.00665$ for
$A=163$, the value $C=0.0075$ has often been used for the TSD
bands in Lu nuclei, see e.g. Refs.~\cite{16106,16302,16407}. 
This larger value of $C$ leads to a substanaial
down-slopes for the observed energies of these bands,
while these energies are rather constant with our standard choice for $C$.   

The absolute value of $E(I) - E_{ref}$  is dependent 
not only on the [shell] energy for a specific spin value but also on 
the [shell] energy at the ground state. This appears reasonable for 
low- and intermediate-spin states formed at similar deformation 
as the ground state. However, for higher spin values, the  deformation or
coupling scheme can be quite different and it is then more reasonable to find
an absolute reference, independent of the ground state for that specific
nucleus. Such an absolute reference is provided by the rotating liquid drop
(RLD) 
model \cite{Coh74}, which can be used in a similar way as a static liquid drop model
is used for 
nuclear ground states \cite{Mye66,Mol95}. 
With this in mind, a RLD reference was introduced in Ref.~\cite{BGC},
where it was concluded that a good fit to  nuclear high-spin states could
be achieved using the Lublin-Strasbourg drop (LSD) model \cite{KP2} for the static
liquid drop energy with the rigid body moment of inertia calculated with
a radius parameter $r_0 = 1.16$ fm and a diffuseness $a=0.6$ fm
\cite{Dav76}. With 
this choice, it becomes possible to describe the absolute energy of
nuclear high-spin state with a similar accuracy ($\sim \pm 1$ MeV)
as nuclear masses \cite{BGC}.   


\begin{figure}[ht]
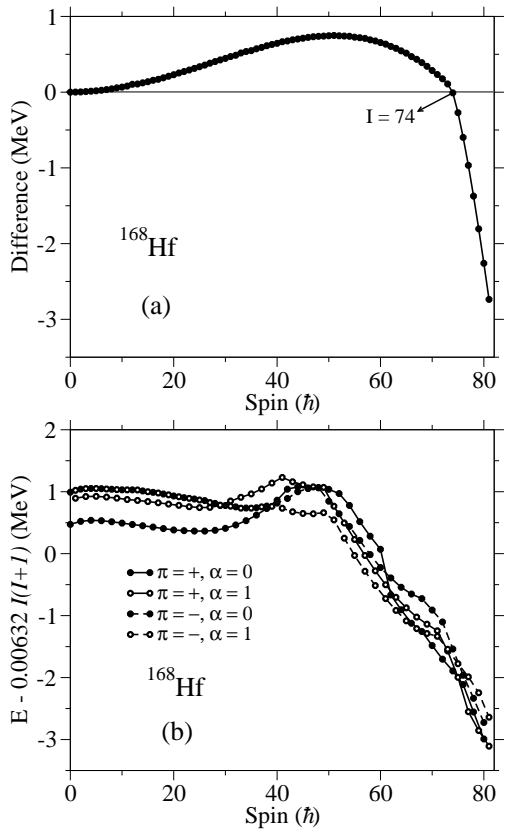

\begin{centering}
\includegraphics[clip=true, scale=0.3]{difference}
\includegraphics[clip=true, scale=0.3]{Hf168-scref}
\caption{\label{refen}{\small (a) 
The difference between the reference energy based on the LSD model with
a moment of inertia calculated from a diffuse surface \cite{BGC},
and the standard $I(I+1)$ reference for the $^{168}$Hf nucleus. 
(b) The $^{168}$Hf yrast energies 
relative to standard 
$I(I+1)$ reference. These energies, which are minimized in the 
$(\varepsilon_2,\gamma,\varepsilon_{4})$ space of deformation,
cf. Fig. \ref{4}, are shown as a function of spin $I$ 
for the four combinations of parity and signature, 
$(+, 0)$, $(+, 1)$, $(-, 0)$ and $(-, 1)$. }}
\end{centering}
\end{figure}
The rotating liquid drop energy at its equilibrium deformation is
plotted relative 
to the fixed reference $C \cdot I(I+1)$ in Fig.~\ref{refen}(a).
This value is thus showing the difference what concerns spin dependence
of the `previous' and 'present' reference energies. Note that both these
references are the same for all bands in one nucleus, but that the mass
dependence is somewhat different. It is easy to understand the general
structure of the curve in Fig.~\ref{refen}(a). At low spin values, 
the equilibrium
deformation of the rotating liquid drop energy is spherical corresponding
to a small moment of inertia and thus a larger reference energy. With
increasing spin, the increasing oblate deformation of the rotating
liquid drop energy corresponds to an increasing rigid body moment
of inertia and at $I \approx 50$, the difference starts to decrease
corresponding to the same moment of inertia for the two reference
energies. At even higher spin at $I \approx 74$, the so-called
superbackbend occurs \cite{Ban75,GA}, when the rotating liquid drop
energy loses its stability towards triaxial shape.
This corresponds to a rapid
increase of the rigid body moment of inertia, 
leading to large negative values for higher
spin values in Fig.~\ref{refen}(a). 

It is now easy to understand
the differences when the yrast energies are plotted relative to
the two differences in Figs.~\ref{4} and \ref{refen}(b), respectively.
Thus the general appearance is the same up to $I = 60-70$ but with
a larger tendency for decreasing values at low spin
with the rotating liquid drop reference. The large differences are however
at the highest spin values where the equilibrium deformations in the
CNS calculations are generally found at a large deformation with
 a small moment of inertia which corresponds a large down-slope
when this energy is shown relative to the $C \cdot I(I+1)$ reference,
see Fig.~\ref{refen}(b). With the rotating liquid drop reference on
the other hand, the reference energies and CNS energies will on the
average have the same spin dependence but a not so nice feature is
that the large changes in the reference energy at the superbackbend
leads to a somewhat strange behaviour of the energies at $I \approx 74$
in Fig.~\ref{4}. 

Let us also point out that the smaller radius parameter combined with the 
diffuseness correction corresponds to essentially the same rigid 
moments of inertia in the two reference energies for mass numbers
$A = 150-200$. For smaller mass numbers on the other hand, the
diffuseness correction becomes more important. For example,
in the $A=60$ region, the spin dependence of the two references
is very similar for spin values $I= 0-15$ but they become quite
different at higher spin values. Thus, already at $I=30$,
the energy of the rotating liquid drop reference is 2-3 MeV 
smaller than the standard $C \cdot I(I+1)$ reference.
  


\section{The high-spin bands in $^{168}$Hf}
\label{168Hf}

\subsection{Observed high-spin bands in $^{168}$Hf}
\label{obs}
\begin{figure*}
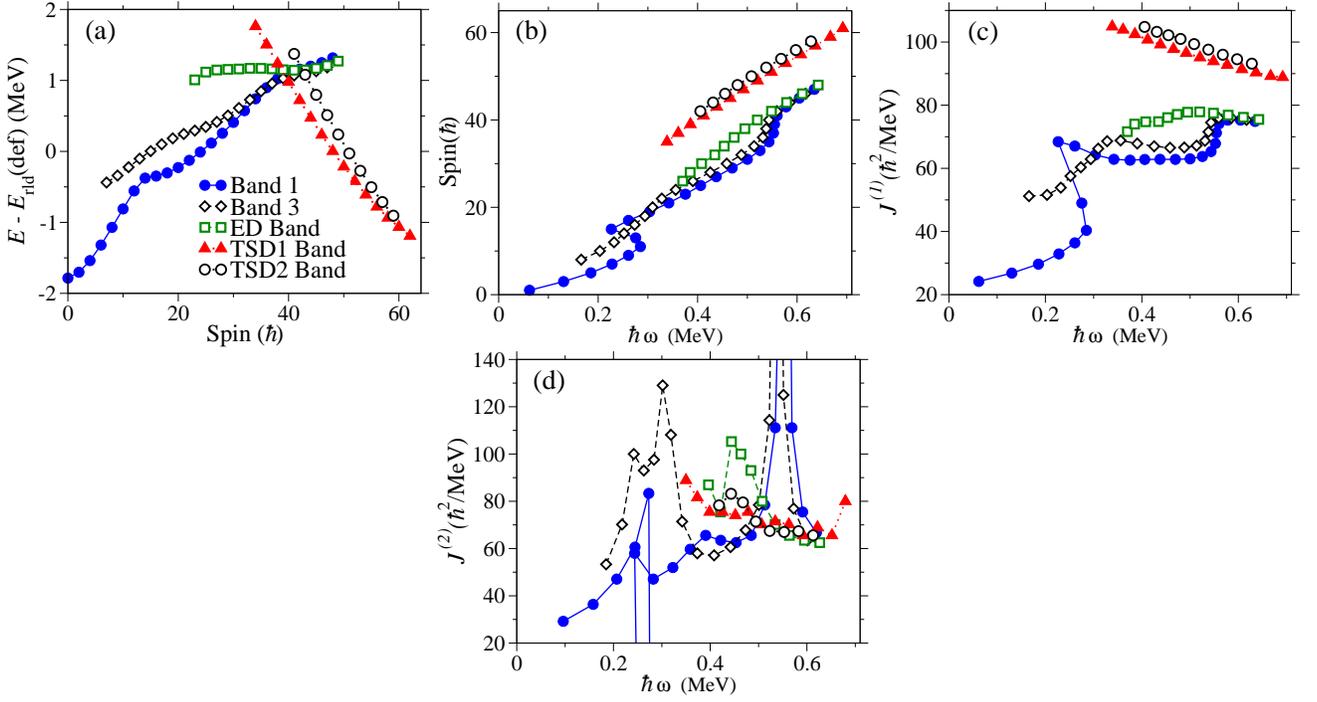

   \centering
\includegraphics[clip=true, scale=0.25]{Hf168-dat}
\includegraphics[clip=true, scale=0.25]{ivseg-Hf168-dat}
\includegraphics[clip=true, scale=0.25]{j1-Hf168-dat}
\includegraphics[clip=true, scale=0.25]{j2-Hf168-dat}
\caption{\label{1}\small{(Color online) (a) Experimental excitation energies relative to that of a
rotating liquid drop E$_{rld}$ as a function of spin $I$, (b) spin
as a function of the rotational frequency, (c) kinematic $({J}^{(1)})$ and (d) dynamic $({J}^{(2)})$
moments of inertia as a function of rotational frequency, $\hbar\omega$ for band 1, band 3, ED, TSD1 and TSD2 bands 
in $^{168}$Hf. The data are taken from Refs.~\cite{RY,RY2,HA2}.}}
\end{figure*}
Experimental excitation energies
relative to a rotating liquid drop energy, E$_{rld}$, as a function
of spin $I$ and spin, kinematic $({J}^{(1)})$ and dynamic $({J}^{(2)})$
moment of inertia as a function of rotational frequency, $\hbar\omega$,
are drawn in Fig.~\ref{1}(a-d), respectively, for the five bands in $^{168}$Hf
which are observed well beyond $I=40$, where pairing correlations should be
negligible. 
From Fig.~\ref{1}(a) one can see that there is a break in the rotational
pattern at $I\sim12$ and $I\sim40$ in band 1 and at $I\sim20$ and
$I\sim40$ in band 3. Furthermore, the spin (Fig.~\ref{1}(b)) and the
${J}^{(1)}$ moment of inertia (Fig.~\ref{1}(c))
are triple-valued for band 1 at $I\sim12$, i.e. band 1 goes through a full
backbend at this spin value. The source of this backbend is the decoupling 
and spin alignment of
an $i_{13/2}$ neutron pair from the pairing field~\cite{Jan81}. 
The unsmoothness in
${J}^{(1)}$ and a small peak in ${J}^{(2)}$ (Fig.~\ref{1}(d)) at $I\sim20$ in band 3
indicates a weak crossing at this spin. The larger variation of
${J}^{(1)}$ and a huge jump in ${J}^{(2)}$ 
at $I\sim40$ ($\hbar\omega\sim0.55$) correspond to a larger
spin alignment in band 1 and band 3 at
this spin. The excitation energy varies smoothly for the ED, TSD1 and TSD2 bands,
which means there is no crossing in these bands, even though the ED band displays
a small rise or bump in the ${J}^{(2)}$ value with the maximum at $\sim0.45$
MeV. 

\subsection{Calculated rotational band structures in $^{168}$Hf}
\label{calc}
\begin{figure}
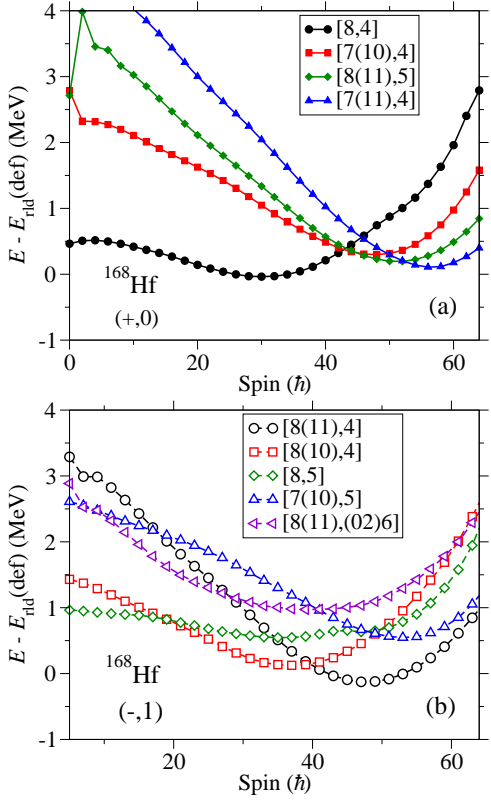

\begin{centering}
\includegraphics[clip=true,scale=0.29]{Hf168-fixpp}
\includegraphics[clip=true,scale=0.29]{Hf168-fixmm}

\caption{\label{conf}{\small{(Color online) Calculated total excitation energies in $^{168}$Hf relative to a rotating liquid drop reference for low-energy configurations with 
(a) $(\pi, \alpha)=(+,0)$ and $\gamma \sim 0^{\circ}$, (b) $(\pi, \alpha)=(-,1)$ and $\gamma \sim 0^{\circ}$.
Each band is shown by a label which is explained in text. 
}}}
\end{centering}
\end{figure}

\begin{figure}[ht]
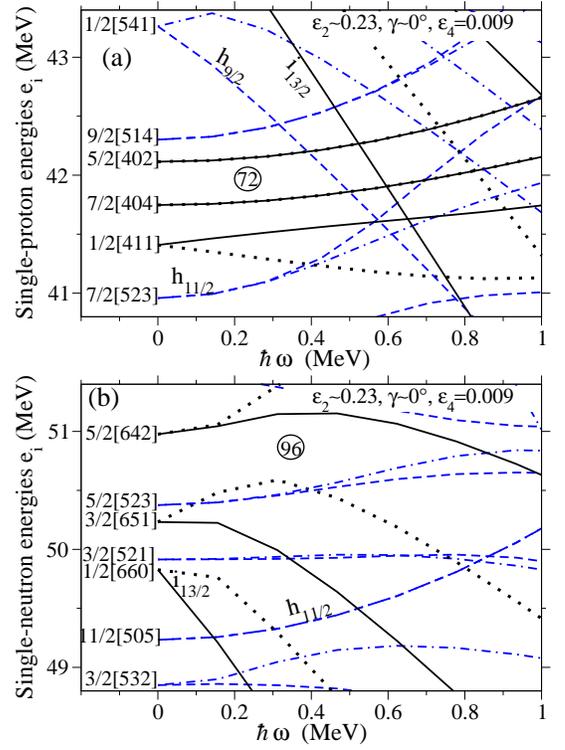

\begin{centering}
\includegraphics[clip=true, scale=0.27]{routh-prsym}
\includegraphics[clip=true, scale=0.27]{routh-nesym}
\caption{\label{8}{\small(Color online) Single-particle proton (a) and neutron (b) energies as a function
of rotational frequency (Routhians) at the deformation $\varepsilon_{2}\sim0.23$,
$\gamma\sim0^{\circ}$ and $\varepsilon_{4}=0.009$. The orbitals
are labeled at $\hbar\omega=0$ by the asymptotic quantum numbers. A few 
important orbitals for the present interpretation are labeled by their dominating $j$-shell.
The line types distinguish between different $(\pi,\alpha)$ combinations:
solid lines represent $(+,+1/2)$, dotted lines $(+,-1/2)$, dashed
lines $(-,+1/2)$ and dash-dotted lines $(-,-1/2)$. }}
\end{centering}
\end{figure}

For the prolate shape minimum (at $\varepsilon_2\sim0.23$) for
$(\pi,\alpha)=(+,0)$ and $(-,1)$, calculated excitation energies for the low-energy configurations
in $^{168}$Hf are plotted, relative to that of a rotating liquid drop
in Figs.~\ref{conf}(a) and (b), respectively. 
As pointed out in section~\ref{min}, these configurations are obtained from
energy minimization in the deformation space
$(\varepsilon_2,\gamma,\varepsilon_4)$.  
The $(+,0)$ yrast line has the configuration
$\pi(h_{11/2})^{8}\nu(i_{13/2})^{4}$ or {[}8,4{]} in the shorthand notation
for spins $I\sim0-40$, see Fig.~\ref{conf}(a). 
As the angular momentum increases, the lowest state is
obtained by exciting a proton from a high-$j$ orbital of $h_{11/2}$ character
to an orbital of $h_{9/2}f_{7/2}$ character in the ${\cal N}=5$
shell. Therefore the yrast line is built from the
$\pi(h_{11/2})^{7}(h_{9/2}f_{7/2})^{1}\nu(i_{13/2})^{4}$ orbitals or {[}7(10),4{]} in a short spin range for $I\gtrsim40$. Then for $I\sim44$,
the calculated yrast configuration is [8(11),5] before the
[7(11),4] configuration comes lowest in energy at $I\sim50$. The single-particle occupancy
in these configurations can be understood from Figs.~\ref{8}(a) and~\ref{8}(b), 
where the single-particle Routhians are plotted
for protons and neutrons,
respectively. The configuration change in the $(+,0)$ yrast states at
$I\sim40$ is
explained from the crossing between the 7/2[523] and 1/2[541] orbitals at
$\hbar\omega\sim0.6$ MeV.
There is a large single-particle
shell gap associated with neutron number $N=96$ 
that continues to $\hbar\omega\sim0.7$ MeV, see Fig.~\ref{8}(b),
so the neutron configuration $\nu(i_{13/2})^{4}$ is favoured up to spin
values beyond $I=50$.  

The study of the calculated excitation energies for the 
low-energy configurations with $(\pi,\alpha)=(-,1)$ and 
axially symmetric shapes (Fig.~\ref{conf}(b))
suggests that the $(-,1)$ yrast line is built on configurations {[}8(10),4{]}
and {[}8(11),4{]} which correspond to $\pi(h_{11/2})^{8}(h_{9/2}f_{7/2})^{1}\nu(i_{13/2})^{4}$
and $\pi(h_{11/2})^{8}(h_{9/2}f_{7/2})^{1}(i_{13/2})^{1}\nu(i_{13/2})^{4}$,
respectively. 
These two bands which cross at $I \sim 40$ have 
the same $(i_{13/2})^{4}$ neutron configuration. The calculated deformations are
 $(\varepsilon_2,\gamma)\sim(0.23,0^{\circ})$ and $(0.26,5^{\circ})$ for {[}8(10),4{]} and {[}8(11),4{]} configurations, respectively.
In Fig.~\ref{conf}(b), also the {[}8,5{]} and {[}7(10),5{]} configurations
are drawn.
They have normal deformation, $(\varepsilon_2,\gamma)\sim(0.23,0^{\circ})$ and cross at spin $I\sim50$. 
The configuration $\pi(h_{11/2})^{8}(h_{9/2}f_{7/2})^{1}(i_{13/2})^{1}\nu(h_{11/2})^{-2}(i_{13/2})^{6}$ or
{[}8(11),(02)6{]} in shorthand notation has the deformation $(\varepsilon_{2},
\gamma)\sim(0.3,1^{\circ})$. 
In fact, the two holes in $h_{11/2}$ neutron orbitals 
(see Fig.~\ref{8}(b)) lead to an enhanced deformation.
\begin{figure}
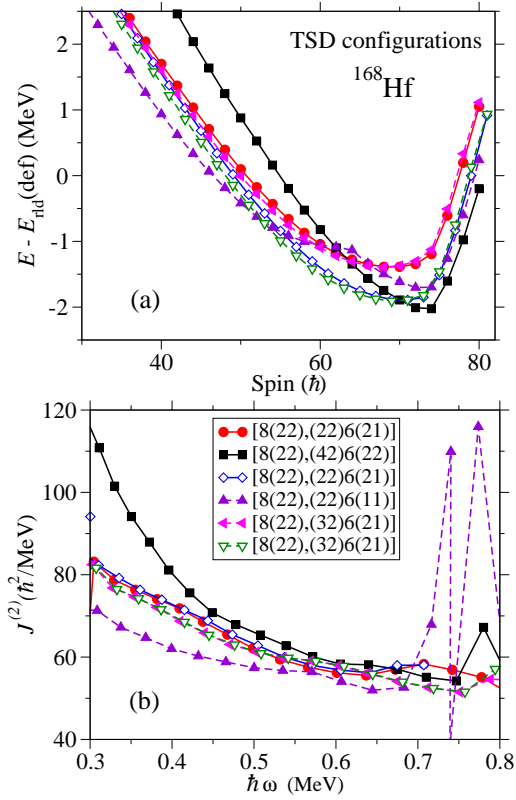

\begin{centering}
\includegraphics[clip=true,scale=0.29]{TSD0}
\includegraphics[clip=true,scale=0.29]{j2-Hf168-cfix}

\caption{\label{conftsd}{\small{(Color online) (a) Calculated total excitation energies relative to a rotating liquid drop reference as a function 
of spin for six low-energy configurations with TSD shape and (b) $J^{(2)}$ values as a function of rotational frequency for the six low-lying 
collective configurations in $^{168}$Hf.
 Each band is shown by a label which is explained in text. Solid lines correspond to positive parity configurations and broken lines correspond to negative parity.
Similarly, solid symbols correspond to signature $\alpha = 0$ and open
symbols correspond to signature $\alpha = 1$.}}}
\end{centering}
\end{figure}

\begin{figure}[ht]
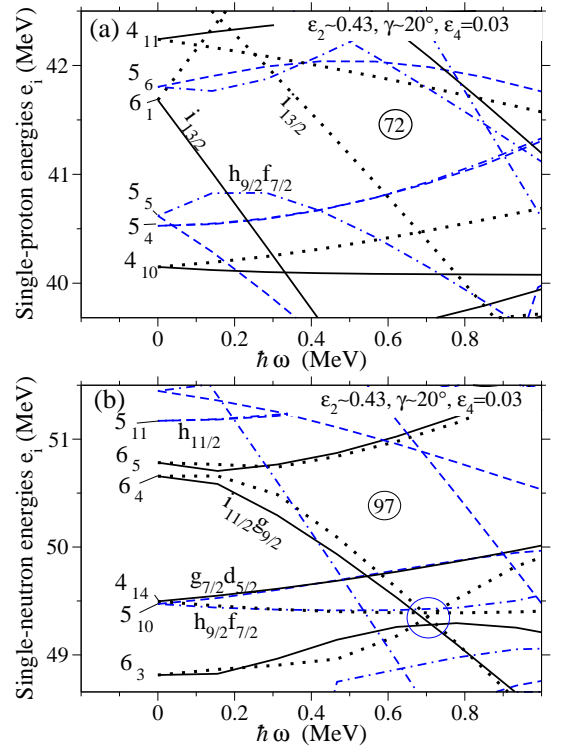

\begin{centering}
\includegraphics[clip=true, scale=0.27]{routh-pr}
\includegraphics[clip=true, scale=0.27]{routh-ne}

\caption{\label{9}{\small(Color online) Single-particle proton (a) and neutron (b) energies as a function
of rotational frequency (Routhians) at the deformation $\varepsilon_{2}\sim0.43$,
$\gamma\sim20^{\circ}$ and $\varepsilon_{4}=0.03$. The orbitals
are labeled at $\hbar\omega=0$ by the ${\cal N}$ shell to which they
belong with the ordering within the ${\cal N}$ shell as a subscript. A few 
important orbitals for the present interpretation are also labeled
by their dominating $j$-shell(s). The line types distinguish
between different $(\pi,\alpha)$ combinations: solid lines represent
$(+,+1/2)$, dotted lines $(+,-1/2)$, dashed lines $(-,+1/2)$ and
dash-dotted lines $(-,-1/2).$ }}

\end{centering}
\end{figure}

 The calculated energies at the TSD minimum are drawn in Fig.~\ref{conftsd}(a)
for six low-energy configurations of $^{168}$Hf. 
The associated dynamic moments of inertia are given as a function of rotational frequency in Fig.~\ref{conftsd}(b).
As discussed above, the total energy is minimized in a five dimensional
deformation space
$(\varepsilon_2,\gamma,\varepsilon_{40},\varepsilon_{42},\varepsilon_{44})$ in
this case.   
All TSD bands are built on the proton configuration $\pi(h_{11/2})^{8}(h_{9/2}f_{7/2})^{2}(i_{13/2})^{2}$
or {[}8(22){]}. This is understood from a proton
single-particle shell gap at $(\varepsilon_{2}, \gamma)\sim(0.43,20^{\circ})$ for $Z=72$ which is seen
in Fig.~\ref{9}(a). 
For neutrons at TSD deformation, a large energy gap is calculated for 
$N=97$ as anticipated from Fig. \ref{spn2} and seen Fig.~\ref{9}(b). 
This suggests
that $^{169}$Hf should be a good candidate to observe TSD bands experimentally.
The lowest $N=96$ configurations are formed from a neutron
hole in the two signatures of the  $(g_{7/2}d_{5/2})$, $(h_{9/2}f_{7/2})$
and $(i_{11/2}g_{9/2})$ orbitals below the $N=97$ gap, 
resulting in six different neutron configurations.
Five of these together with
one configuration with two ${\cal N}=7$ neutrons are
\begin{figure}
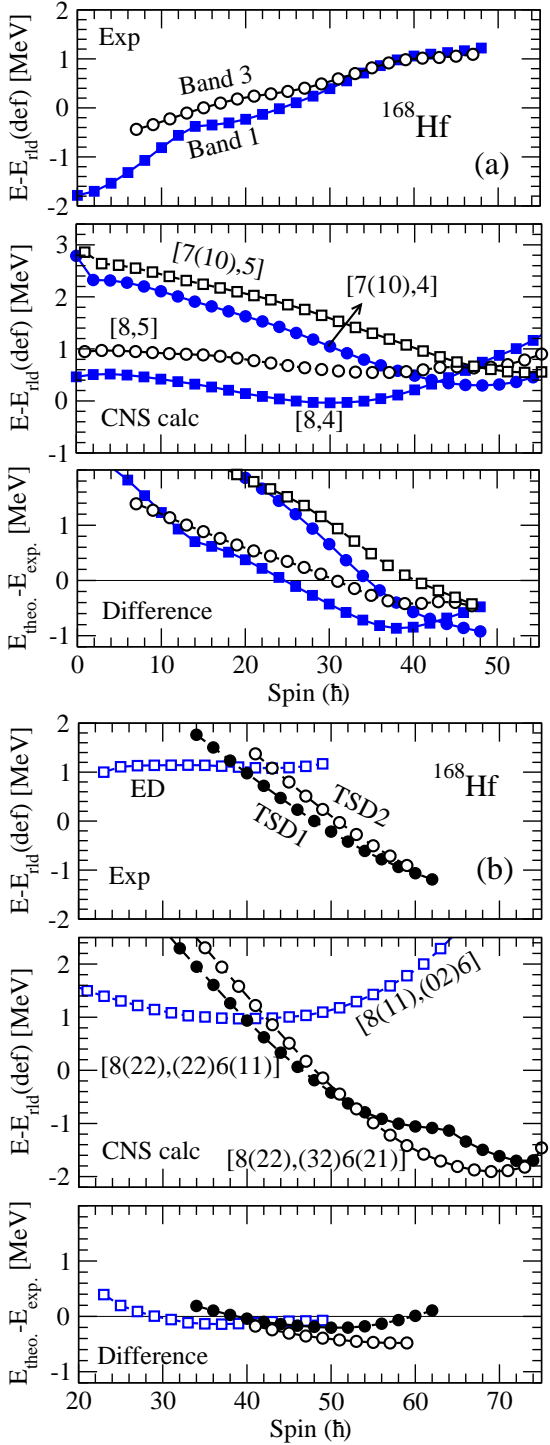

\begin{centering}
\includegraphics[clip=true,scale=0.57]{expthg3m}
\includegraphics[clip=true,scale=0.57]{expthEDTm}
\caption{\label{diff}{\small(Color online) Experimental energies (top panels) and theoretical energies (middle
panels) relative to a rotating liquid drop and their differences (lower panels) as a function of spin
 for (a) band 1 and band 3 and (b) ED, TSD1 and TSD2 bands in $^{168}$Hf.}}
\end{centering}
\end{figure}
combined with the favored
proton configuration, forming six the low-energy triaxial structures 
shown in Fig.~\ref{conftsd}(a). Note that
all the neutron configurations are built on six neutrons in $i_{13/2}$ orbitals, two holes in
$h_{11/2}$ orbitals and two, three or four holes in ${\cal N}=4$ orbitals. 
The calculated deformation is $(\varepsilon_2,\gamma)\sim(0.49,22^{\circ})$ for
the {[}8(22),(42)6(22){]} 
configuration while it is $(\varepsilon_2,\gamma)\sim(0.4-0.45,20^{\circ})$ for
the other TSD configurations. 
All of the theoretical TSD bands shown in Fig.~\ref{conftsd}(a) display a decreasing value of $J^{(2)}$ 
with increasing rotational frequency. At $\hbar\omega\lesssim0.45$ MeV, the value of $J^{(2)}$ decreases more strongly for the configuration 
[8(22),(42)6(22)] and more smoothly for the [8(22),(22)6(11)]. The values of $J^{(2)}$ are very close together at $\hbar\omega\gtrsim0.45$ MeV and only 
the configuration [8(22),(22)6(11)] experiences a sharp discontinuity in the $J^{(2)}$ moment of inertia at $\hbar\omega\sim0.7-0.8$ MeV. This discontinuity 
is because of a crossing
in the neutron single particle orbitals between a high-$j$
$i_{13/2}$ orbital and a low-$j$ $(i_{11/2}g_{9/2})$ orbital 
at $\hbar\omega\sim0.7$ MeV. The crossing is indicated by a circle in Fig.~\ref{9}(b). 
In the other TSD configurations the $(i_{11/2}g_{9/2})$ orbital has been filled and therefore the strong alignment 
at $\hbar\omega\sim0.7$ MeV is blocked and there is no anomaly in the $J^{(2)}$ moment of inertia of them.

 In the calculations, no distinction is made
between low-$j$ and high-$j$ orbitals at this large deformation, i.e.
only the number of particles of signature $\alpha = 1/2$ and 
$\alpha = -1/2$ in each ${\cal N}_{rot}$-shell is fixed. 
On the other hand, the configurations are labeled as if such a
distinction is made. The labels in Fig.~\ref{conftsd}(a) refer to the
configuration for spin values below $I\sim60$. For example,
the energy of the  configuration labeled {[}8(22),(22)6(11){]} comes down at spin
$I\sim62$, because of the crossing discussed above. Thus, the band should be labeled
{[}8(22),(22)5(21){]} for higher spin values.

\subsection{Comparison between calculated and experimental bands in $^{168}$Hf}
\label{comp}

In the upper panels of Fig.~\ref{diff}, experimental excitation energies relative 
to a rotating liquid drop energy for band 1, band 3, the ED band and the TSD1 and TSD2  
bands are drawn as a function of spin. The middle panels of Fig.~\ref{diff} displays the 
calculated bands which seem to be closest to these experimental bands. 
In the lower panels, experimental and theoretical bands are compared 
(in attention to their parity and signature) and their differences are illustrated. 

\subsubsection{Band 1}
The observed band 1 has a positive parity and signature $\alpha=0$. Therefore
one can find out the structure of this band from the search among the
lowest-energy configurations which have $(\pi,\alpha)=(+,0)$
(Fig.~\ref{conf}(a)). As it is pointed in section~\ref{calc}, in
Fig.~\ref{conf}(a) the {[}8,4{]} and {[}7(10),4{]} configurations are the
lowest states in energy at spins $I\leq40$ and $I\geq40$, respectively.  The
{[}8,4{]} configuration has an even number of neutrons in $i_{13/2}$
orbitals. Thus the observed backbending at $I\sim12$ in band 1 (see section~
\ref{obs}) could occur in this configuration. Furthermore, the change of structure
from {[}8,4{]} to {[}7(10),4{]} which happens at spin $I\sim40$
corresponds to the observed break in the rotational pattern at $I\sim40$ in
this band (see Figs.~\ref{1}(a-d)). Band 1 is compared with the {[}8,4{]} and
{[}7(10),4{]} configurations in the lower panel of Fig.~\ref{diff}(a). As one
can see the differences between the theoretical and experimental data for band
1 is rather constant and about $-1$ MeV at spins $I\geq35$ if a transition
occurs from {[}8,4{]} to {[}7(10),4{]}.  

The same structure 
has been
obtained for band 1 in Ref.~\cite{RY2} using the Ultimate Cranker code~\cite{RBc,TB90}. 
However, it is concluded~\cite{RY2} that the occupation of the 1/2 [541]
orbital at $I\sim40$
is related to the crossing between the
proton orbitals 9/2[514] and 1/2[541]. 
This is contrary to our calculations, see
Fig.~\ref{8}(a), where the 9/2[514] orbital is above the Fermi
surface and thus the transition is because of the crossing between the
7/2[523] and 1/2[541] orbitals. A closer look at Fig. 5 of Ref.~\cite{RY2}
indicates that there are two $h_{11/2}$ quasiparticles at similar energies
where one should then mainly correspond to a hole in the  7/2[523]
orbital and the other to  a particle in the 9/2[514] orbital.
Then, it appears that if the number of particles should 
not be changed drastically, the
added particle in 1/2[541] should be combined with a hole in 7/2[523],
contrary to the conclusion in Ref.~\cite{RY2}.

\subsubsection{Band 3}
Band 3 is observed from $I=7$ to $I=47$ and has $(\pi,\alpha)=(-,1)$.
Excitation energies, spin, ${J}^{(1)}$ and ${J}^{(2)}$ behavior of
this band are close to those of band 1 (see Figs.~\ref{1}(a-d)). Thus it seems
that this band has a similar deformation as band 1. The lowest-energy states
with $(\pi,\alpha)=(-,1)$, see Fig.~\ref{conf}(b), suggest 
that the {[}8(10),4{]} and
{[}8(11),4{]} configurations should be assigned to band 3. 
However, as we see in Figs.~\ref{1}(a-c), in
contrast to band 1, band 3 does not backbend at low spins. Therefore the
neutron configuration of this band could not be the same as that of band 1 
(which has four neutrons in 
$i_{13/2}$). As pointed in section~\ref{calc}, the {[}8,5{]} and
{[}7(10),5{]} configurations have almost the same deformation as band 1 and
they also have an odd number of neutrons in $i_{13/2}$ orbitals. Thus,
even though these two configurations are calculated about 0.5 MeV higher in energy than
{[}8(10),4{]} and {[}8(11),4{]}, see Fig.~\ref{conf}(b), they are more
suitable candidates for band 3. As one can see in Fig.~\ref{conf}(b), there is
a crossing between {[}8,5{]} and {[}7(10),5{]} at spin $I\sim50$ which is in
agreement with the observed crossing in band 3 experimentally.  In the lower
panel of Fig.~\ref{diff}(a) band 3 has been compared to the {[}8,5{]} and
{[}7(10),5{]} configurations. With the transition from the {[}8,5{]} to
the {[}7(10),5{]} configuration, the differences between the calculated bands and band 3 is almost
constant at -0.5 MeV for spins $I\geq35$.  Therefore band 3 is built
from the ${(i_{13/2})}^5$ neutron configuration and the proton configuration
is the same as that for band 1, ${(h_{11/2})}^8$ at $I\lesssim50$ and
${(h_{11/2})}^7{(h_{9/2}f_{7/2})}^1$ for $I\gtrsim50$. This interpretation
is similar to that  of Ref.~\cite{RY2}, but with the same difference
as discussed above for band 1. 

\subsubsection{Band ED}
The ED band (called TSD2 in Ref.~\cite{HA2}) has $(\pi,\alpha)=(-,1)$ and is 
observed from $I=23$ to $I=49$. The calculations, as depicted in Fig.~\ref{conf}(b), show that
 $\pi(h_{11/2})^{8}(h_{9/2}f_{7/2})^{1}(i_{13/2})^{1}\nu(i_{13/2})^{4}$ or the {[}8(11),4{]} configuration is lowest
 in energy for spins $I\gtrsim40$ 
This configuration, which corresponds to axially symmetric shape,
has been suggested for the ED band in $^{168}$Hf~\cite{RY}.
The calculated quadrupole deformation value,  $\varepsilon_2 \sim 0.26, \gamma
\sim 5^{\circ}$,
is near normal deformation and far from that of the ED bands in the other Hf
isotopes, $\varepsilon_2\sim0.3$~\cite{YZ}.
The suggested configurations for the ED band in Hf isotopes are all built on the same proton configuration, $i_{13/2}h_{9/2}$, but they are coupled 
to different neutron configurations~\cite{YZ,RY}. 

A common feature of most interpretations of strongly collective bands is that only the
high-$j$ intruder orbitals from the higher shells are listed explicitly in
the configurations. 
These high-$j$ orbitals are important to build the spin
but on the other hand, it is rather the extruder orbitals from the lower
shells which build the collectivity. This is evident for the smooth
terminating bands~\cite{AA} and it has been underlined that it is the case
for more collective bands, see e.g.~\cite{Pau07, Rag96}. However, to our
knowledge, it is only in the present CNS approach that methods have been
developed to fix the number of particles in the extruder orbitals.
In the present case, the highest $h_{11/2}$ orbital is just below the
Fermi surface. Thus we consider the configuration with two holes
in the upsloping 11/2[505] orbital, i.e. {[}8(11),(02)6{]}.
This configuration is about 1 MeV higher than {[}8(11),4{]} in energy, 
see Fig.~\ref{conf}(b). However, 
as pointed in section~\ref{calc}, the {[}8(11),(02)6{]} configuration has  
a larger value for the 
quadrupole deformation, $(\varepsilon_{2},\gamma)\sim(0.3,1^{\circ})$, which is in
agreement with that of in the other Hf isotopes. Especially, the experimental properties of the
ED band are clearly different from those of the valence space band and it
is only with the holes in the  $h_{11/2}$ neutron orbitals that also the
theoretical configuration becomes clearly different from the valence
space configurations.  
With the {[}8(11),(02)6{]} interpretation for the ED band, the difference
between calculations and experiment becomes small and almost constant
as seen in the lower panel of Fig.~\ref{diff}(b). It also suggests that 
such configurations with holes on the $h_{11/2}$ neutron orbital, below $N=82$, should be investigated 
for the ED bands in  other Hf isotopes. Furthermore, as mentioned in
Ref.~\cite{Pau07}, it appears that the same mechanism with holes in
the $h_{11/2}$ neutron orbitals is responsible for the large quadrupole
moment in the SD band of $^{175}$Hf.


One problem with the present interpretation is that the {[}8(11),(02)6{]
configuration is calculated at an excitation energy which is somewhat too
high relative to the configurations assigned to band 3. However, these
differences are clearly within the expected uncertainties. 
For example, if the neutron $h_{11/2}$ subshell was placed 0.5 MeV higher
in energy, the {[}8(11),(02)6{] configuration would be calculated
close to yrast.  Note also that
when the {[}8(11),(02)6{] configuration is not calculated as yrast, 
it is straightforward to 
study it only in approaches like the present one where it is possible to fix
the number of holes (or particles) in specific orbitals.

\subsubsection{TSD1 and TSD2 Bands}
\label{TSD}   
The observed TSD bands in $^{168}$Hf~\cite{RY} have not been linked
to the normal-deformed level scheme so their spin and excitation
energy can only be estimated.
The structure $\pi(i_{13/2})^{2}\nu(j_{15/2}i_{13/2})$ has been suggested as
the most probable intrinsic configuration for the 
TSD1 band.
Our calculations with a complete minimization 
in $\varepsilon_4$ show that
the {[}8(22),(22)6(11){]} configuration is the lowest-energy TSD configuration
at spins $I\lesssim50$, see Fig.~\ref{conftsd}(a). Note that these two
suggested
configurations are identical with the same high-$j$ orbitals occupied but,
in the unpaired CNS formalism,
also the occupation of other orbitals are specified, including the extruder
neutron orbitals with their main amplitudes in the $h_{11/2}$ and ${\cal N}=4$
shells, respectively.

The {[}8(22),(22)6(11){]} 
configuration with $(\pi,\alpha)=(-,0)$ is about 0.5 MeV lower than the next
lowest-energy TSD configuration ({[}8(22),(32)6(21){]} with $\alpha=1$)
at spins $I\lesssim45$.  Band TSD1 has been
measured up to $I\sim60$ where the {[}8(22),(22)6(11){]} configuration is
calculated only a few hundred keV above the yrast line. Thus, we choose this 
configuration as our favored candidate for TSD1 and compare the two bands in
lower panel of Fig.~\ref{diff}(b). 
This configuration choice 
suggests that the observed 
band TSD1 has negative parity and even spin ($\alpha=0$).


TSD1 is plotted with an assumed bandhead spin of
$I=34\hbar$ 
and with an energy of 11.6 MeV for the bandhead.
This leads to a good fit between the observed band and  
the calculated configuration
{[}8(22),(22)6(11){]}, where the shorthand notation corresponds to
$\pi(h_{11/2})^{8}(h_{9/2}f_{7/2})^{2}(i_{13/2})^{2}$ and 
$\nu({\cal N}=4)^{-2}(h_{11/2})^{-2}(i_{13/2})^{6}(i_{11/2}g_{9/2})^1(j_{15/2})^1$
for the occupation of the open proton and neutron subshells. 
The calculated quadrupole
moment of the configuration {[}8(22),(22)6(11){]} is in the range 11.8-9.9 eb for
spin values $I=30-80$ which is in agreement with the experimentally measured
value of $Q_t=11.4_{-1.2}^{+1.1}$ eb~\cite{HA2}. However, the calculated
quadrupole moment is similar for all configurations in the TSD minimum 
so it does not help to  discriminate between the different configurations
listed above.

The configuration assignment to the TSD1 band is certainly preliminary
and one might for example argue that we should rather choose a configuration
which is calculated yrast at the highest observed spin value, $I\sim60$.
This would then rather suggest [8(22),(22)6(21)] or [8(22),(32)6(21)]
as the favoured choice,  
i.e. configurations with one more neutron excited to the $i_{11/2}g_{9/2}$ 
orbitals. In addition to the high-$j$ particles, the TSD minimum is
characterized by at least 2 $h_{11/2}$ and 2 ${\cal N}=4$ neutron holes.
This is the case also for the calculations presented in Ref.~\cite{RY} even though
these holes are not specified in the configuration labels used in that
reference. As pointed out in section~\ref{Lu}, these holes are important
to create the smooth collective bands where it is mainly the ${\cal N}=4$ holes which
induces the triaxial shape.

Based on comparisons with calculations and with the TSD1 band, see
Figs.~\ref{conftsd}(a) and \ref{diff}(b), the
spin and bandhead energy for the TSD2 band are estimated to 
be $41\hbar$ and 14.7 MeV, 
respectively. 
As pointed out above, the next lowest TSD 
configuration for $I \sim 40-50$ is {[}8(22),(32)6(21){]}  
with $(\pi,\alpha)=(-,1)$. This configuration is yrast 
for $55\lesssim I\lesssim70$. 
Therefore it seems that this configuration 
is a reasonable candidate for the observed TSD2 band.
As one can see in the lower panel of 
Fig.~\ref{diff}(b), the energy difference between the TSD2 band and 
the {[}8(22),(32)6(21){]} configuration 
is small and rather constant for $I\gtrsim50$.    
Considering the configurations in Fig.~\ref{conftsd}(a),
another possible choice is the {[}8(22),(22)6(21){]} 
configuration with $(\pi,\alpha)=(+,1)$.
Thus, the present calculations suggest that compared with TSD1, 
TSD2 has the same proton configuration but with one neutron 
excited to  $i_{11/2}g_{9/2}$ from either the ${\cal N}=4$ orbitals or from
the $h_{9/2}f_{7/2}$ orbitals. This leads to odd spin values
($\alpha =1$) but undetermined parity for TSD2. 


\begin{figure}
\begin{centering}
\includegraphics[clip=true,scale=0.32]{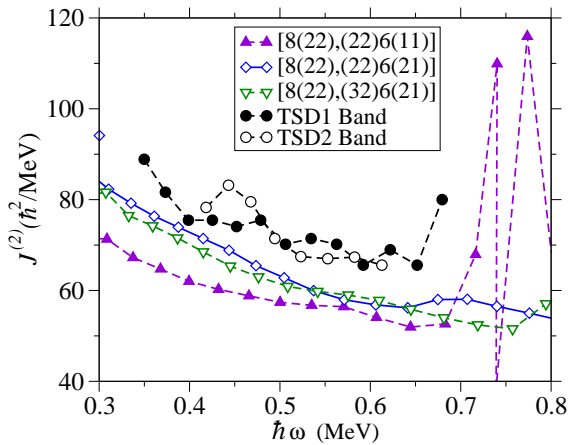}
\caption{\label{j2}{\small(Color online) Theoretical and experimental dynamic moments of inertia $J^{(2)}$ as a function of
 rotation frequency $\hbar\omega$ for $^{168}$Hf.}}

\end{centering}
\end{figure}
 The configurations of the TSD1 and TSD2 bands could be interpreted by considering the behavior of the dynamic 
moment of inertia. Although the spin assignments for these TSD bands may need revising, the dynamic moments of inertia are not 
affected by these changes. A smooth decrease in the $J^{(2)}$ moment is
observed for the TSD1 and TSD2 bands, see Fig.~\ref{j2}, 
which is consistent with the general trend of TSD bands in the other Hf isotopes~\cite{MD,DS,Neu02}.  
Fig.~\ref{j2} also displays three configurations that have characteristics
similar to the two observed TSD bands. On the other hand, the absolute value of $J^{(2)}$
is somewhat smaller in calculations than in experiment, which can also be
concluded from the positive curvature in the difference curves in the lower
panel of Fig. \ref{diff}(b). 
The TSD1 band and the [8(22),(22)6(11)] configuration have rather similar slopes in $J^{(2)}$ throughout the observed frequency 
range. The value of $J^{(2)}$ for the TSD2 band has a behavior similar to that of two suggested configurations but the calculated dynamic 
moment of inertia is the same for the others TSD configurations, see Fig.~\ref{conftsd}(b), so it does not help to
choose a favorable configuration for the TSD2 band.


\section{Additional comments on the TSD bands in $A=158-168$ nuclei.}
\label{Er}

\subsection{Full minimization in Lu isotopes and $^{158}$Er}
We have examined the full minimization approach in the hexadecapole
deformation space (see sect.~\ref{min}) for TSD configurations in Lu isotopes. 
Our calculations show that this effect will typically decrease the minimum energy by
200 keV in the observed spin region ($I\lesssim50$). The maximum gain 
of about 300 keV is obtained 
in $^{165}$Lu, which has a large hexadecapole equilibrium deformation,
$\varepsilon_4\sim0.04$. These effects will lead to some minor 
corrections on the results presented in
sect.~\ref{Lu} but they will clearly not change the general
conclusions.
  
Similar calculations have been carried out for $^{158}$Er where, according 
to studies in Ref.~\cite{Oll09,Wan11}, three well-defined TSD minima with 
deformations $(\varepsilon_{2},\gamma)\sim(0.37,\pm20^{\circ})$ and
$(\varepsilon_{2},\gamma)\sim(0.45,20^{\circ})$ are seen. Our calculations show
that a complete minimization in $\varepsilon_4$ has only a small 
influence on the energy. 
The gain in energy is always smaller than 200 keV for the
TSD configurations in $^{158}$Er.

\subsection{Effective alignments in a larger mass range from $^{158}$Er to $^{168}$Hf}

\begin{figure}
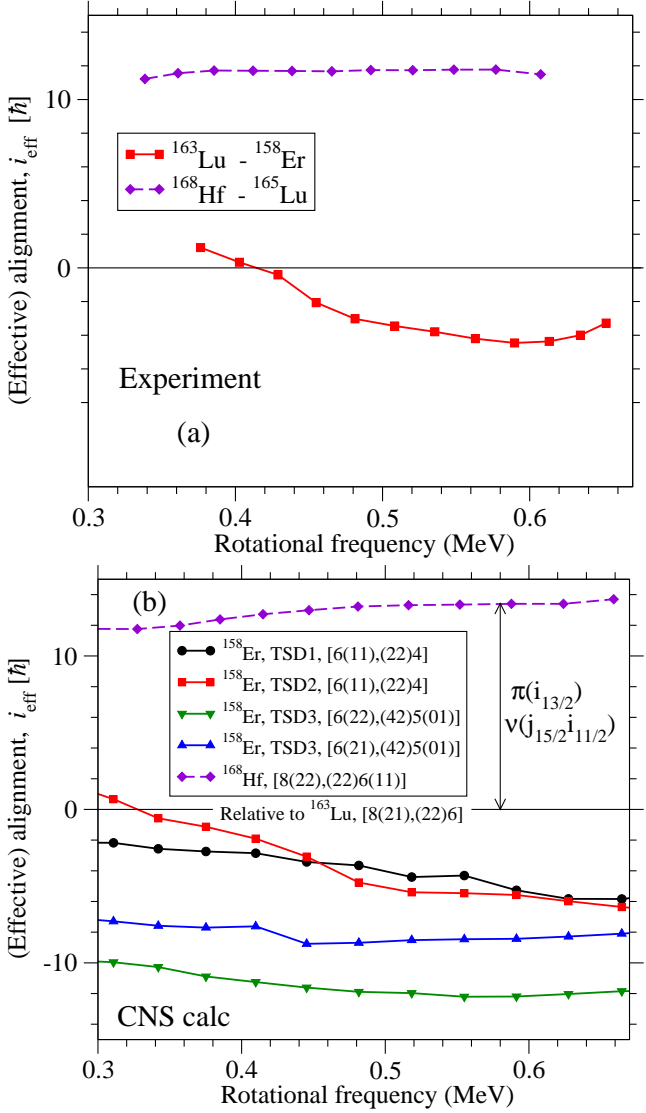

\begin{centering}
\includegraphics[width=8.5cm,clip=true]{ieff-exp-erluhf}
\includegraphics[width=8.5cm,clip=true]{ieff-theo-erluhf}
\caption{\label{ieff2}{\small(Color online) Effective alignment, $i_{eff}$,
extracted from 
(a) experiment and (b) calculated configurations. The TSD bands in $^{163}$Lu and $^{158}$Er, and $^{168}$Hf and
$^{163}$Lu, respectively, are compared when calculating $i_{eff}$.
For $^{158}$Er, configurations at positive $\gamma$ and smaller (TSD1) and
larger (TSD3) $\varepsilon_2$ and negative $\gamma$ (TSD2) are considered.}}
\end{centering}
\end{figure}
It is instructive to consider the alignment in a larger mass range outside the
Lu isotopes. Therefore, in Fig.~\ref{ieff2} the difference in alignment
between the lowest TSD bands in $^{163}$Lu and $^{168}$Hf is shown for the
experimental bands and for the configurations we have assigned to these
bands. In the calculations, we have chosen the spin values for the unlinked
band in $^{168}$Hf based on the present calculations and this will naturally lead to
a general agreement between  $^{163}$Lu and $^{168}$Hf what concerns the 
effective alignment. Considering the orbitals which are filled in $^{168}$Hf
but not in  $^{163}$Lu, it is mainly the $i_{13/2}$ proton and the $j_{15/2}$
and $i_{11/2}$ neutrons which build the large effective alignment of almost
$12\; \hbar$ while the filling of two ${\cal N} = 5$ neutron 
orbitals will only have a small contribution to $i_{eff}$.

It is then also instructive to compare the spin difference between the lowest
TSD bands in  $^{158}$Er and $^{163}$Lu which is drawn for experiment and
calculations in the upper and lower panels of Fig.~\ref{ieff2}, respectively.
Several possible theoretical assignments are shown, corresponding to the
lowest-energy configurations in the minima with $\varepsilon_2 \approx 0.34,
\gamma \approx 20^{\circ}$ (TSD1), $\varepsilon_2 \approx 0.34,
\gamma \approx -20^{\circ}$ (TSD2) and $\varepsilon_2 \approx 0.43,
\gamma \approx 25^{\circ}$ (TSD3), where the different minima are labelled
as in Ref.~\cite{Wan11}. Furthermore, it should be  
noted that the spin values in $^{158}$Er are not known and have been
chosen in the range $I=23-65 \hbar$ 
as suggested in Ref.~\cite{Pau07}. 
If these spin values
are increased (decreased) by $\Delta I\; \hbar$, it will correspond 
to a constant decrease (increase) for values 
of the curve in the upper panel by $\Delta I\; \hbar$, but with no change of the
spin dependence. 

The TSD1 configuration of $^{158}$Er and the TSD band of
$^{163}$Lu have similar deformations so the corresponding value
of $i_{eff}$ measures the spin contribution from the orbitals which are
filled in $^{163}$Lu but not in $^{158}$Er, i.e. 2 $h_{11/2}$ and one $h_{9/2}$ protons
and 2 $i_{13/2}$ neutrons. They will then give a negative contribution to $i_{eff}$
in agreement with general 
expectations for orbitals in the middle of a $j$-shell, see
e.g. Fig. 27 of Ref.~\cite{TB}.
When it
comes to the configurations in the TSD2 and TSD3 minima, 
the value of $i_{eff}$ does not correspond to the contribution of any specific
orbitals because these configurations do not have a common core with
$^{163}$Lu.

In any case, it is still possible to define the difference
in spin value for a fixed frequency and the comparison between experiment
and calculations in Fig.~\ref{ieff2} shows that it is necessary to increase
the spin values in $^{158}$Er by $4-8 \hbar$ to get agreement at the highest
frequencies for the TSD3 configurations. 
For the TSD2 configuration on the other hand,
experiment and calculations come close for all
frequencies with present spin values. Note especially that a down-slope is seen
both in experiment and calculations for frequencies 
$\hbar \omega \approx 0.35 - 0.50$ MeV. This down-slope corresponds to an
additional alignment of $\sim 4 \hbar$ in 
this frequency range for $^{158}$Er relative to $^{163}$Lu.
Such an alignment gives rise to a bump in the ${J}^{(2)}$ moment of
inertia as discussed in some detail for the corresponding bands in 
$^{159,160}$Er in Ref.~\cite{Oll09}. As discussed there, the alignment is
caused by a crossing between an $h_{11/2}$ proton orbital and the lowest
$h_{9/2}$ orbital. As seen in Fig.~\ref{ieff2}, 
while the observed alignment is
approximately reproduced for the TSD2 configuration, no
similar alignment is seen in the TSD1 and TSD3 configurations. In Ref.
\cite{Oll09}, a bump in the ${J}^{(2)}$ moment of inertia for the
TSD1 band, caused by a crossing between the ${\cal N} = 6$ neutron orbitals, was 
discussed. This is however a considerably broader bump corresponding 
to an alignment in a larger frequency range which is not seen as any
well-defined alignment in Fig.~\ref{ieff2}. For the TSD3 configuration,
no specific alignment is seen in  Fig.~\ref{ieff2} and no crossing
between orbitals is observed in Fig.~\ref{9} which could give raise
to such an alignment.

The conclusion from the present analysis of the alignments would 
thus be that the
TSD2 configuration, i.e. the $\gamma < 0$ configuration, which is
the preferred assignment for the yrast TSD band in $^{158}$Er. This is
also the preferred configuration when comparing the transitional
quadrupole moment $Q_t$ \cite{Wan11}, even though the value of
$Q_t$ for the TSD3 configuration is not much different. However,
if a TSD3 configuration is assigned, it  appears to  correspond 
to an unrealistic increase of the spin values in the band 
compared with the values which appears most realistic from an
experimental point of view \cite{Pau07}. The assignment of
a negative $\gamma$ configuration is in disagreement with
Ref.~\cite{Shi12} where it is concluded that a TSD2 minimum
would be instable towards the TSD1 minimum in the tilted axis
degree of freedom. This conclusion however requires that the
TSD1 minimum is (considerably) lower in energy than the TSD2
minimum which is the outcome of the CNS calculations as
well as the calculations of Ref.~\cite{Shi12}. However, the
relative energies of the three TSD minima are clearly uncertain
within at least $\pm 1$ MeV. Thus, considering only the
calculated energies,
it could be a configuration in any of the three TSD
minima which should be assigned to observed yrast TSD 
band in $^{158}$Er. One may also note that with present
interpretations, there is no strong relation between
the TSD bands in $^{158}$Er and $^{168}$Hf.

\
\section{Conclusions}
\label{conclu}

In this study, we have made some modifications when solving the Hamiltonian
in configuration-dependent cranked Nilsson-Strutinsky formalism in order
to test the accuracy of some  approximations. The nucleus $^{168}$Hf is
used as a test case and the observed highest-spin bands in this nucleus 
are analyzed.

An important feature of the CNS formalism is that the off-shell matrix
elements in the rotating harmonic oscillator basis are neglected in order
to make it possible to fix configurations in more detail. 
For the yrast states, it is however straightforward to include all couplings
and compare with the approximate CNS results. We conclude that  
the neglect of the off-shell elements of
the Hamiltonian matrix is acceptable; i.e. 
for the deformations and rotational frequencies which 
are reached in $^{168}$Hf, it only leads to small errors
which  are essentially negligible compared with other uncertainties.  
This is also true for the cut-off error due to the limited number of
oscillator shells in the basis, where our calculations show that
if more than 9 shells (${\cal N}_{max}>8$) are included,
both the total discrete energy and the total smoothed
energy decrease, which means that the shell energy and thus 
also the total rotational energy is almost unchanged.  

The total energy was minimized in the full hexadecapole space indicating that
for axially symmetric shape, it is generally sufficient to include  
only the standard $\varepsilon_{40}$ degree of freedom which does not break the axial
symmetry. On the other hand, at large triaxial deformation, it appears 
necessary to minimize the energy in all three hexadecapole deformation
parameters, $(\varepsilon_{40},\varepsilon_{42},\varepsilon_{44})$.
The study of the deformation space
shows that in $^{168}$Hf, the energy of the axially
symmetric bands with normal deformation could as well be minimized in the
restricted
deformation space $(\varepsilon_2,\gamma,\varepsilon_4)$, while the energy of the
TSD bands should be minimized in the deformation space
$(\varepsilon_2,\gamma,\varepsilon_{40},\varepsilon_{42},\varepsilon_{44})$. We
used these results and studied the structure of the experimental bands band 1, band
3, band ED and bands TSD1 and TSD2 in $^{168}$Hf.

In our studies of the high-spin bands of $^{168}$Hf, the general conclusions
are the same as in Refs.~\cite{RY,RY2} but still with some important
differences. Thus, we conclude that the crossing observed around $I=40$ 
in the normal deformed bands are created when a particle is excited to
the down-sloping 1/2[541] orbital but rather
from the 7/2[523] orbital
and not from the 9/2[514] orbital suggested in Ref.~\cite{RY2}.
The different conclusions are understood from the
fact that the number of particles is preserved in the unpaired formalism,
while
quasi-particle excitations with a fixed Fermi energy~\cite{RY2} could lead to
significant changes in the number of particles. More important is however our
suggestion that the ED band is built with two holes in the extruder
$h_{11/2}$ orbitals. With  such an excitation, the configuration is clearly different
from that of the normal deformed bands with a calculated transitional
quadrupole moment in closer agreement with experiment. Because this
configuration is not calculated as yrast, it is straightforward to
analyze it only in formalisms where excited configurations with the 
same quantum numbers can be distinguished, e.g. by the
number of particles (or holes) in high-$j$ orbitals. For the TSD1 and
TSD2 bands, we conclude
that they are formed in 
a strongly deformed triaxial minimum with several particles excited to
high-$j$ intruder orbitals. This agrees with the assignment in
Ref.~\cite{RY} what concerns TSD1.
However, while only the high-$j$ particles were considered in
that reference, the holes
in the extruder orbitals are as important
according to our analyses, where it is mainly the neutron
${\cal N}=4$ orbitals which induces the triaxial shape.

As a background for the study of the Hf bands, we did also investigate
the filling of the orbitals of the TSD bands in Lu isotopes. These
bands are naturally understood as having a proton configuration with
one $i_{13/2}$ orbital occupied, compared with two such orbitals
filled in the larger deformation TSD bands of $^{168}$Hf. The 
neutron configurations
in the $N=90-96$ range are characterized by a successive filling
of down-sloping orbitals in a region of low level-density which
is created below  four extruder orbitals emerging from the
$h_{11/2}$ and ${\cal N} = 4$ shells, see Figs. \ref{spn} and 
\ref{spn2}.

The relative alignments, $i_{eff}$ \cite{Rag93}, between the TSD bands 
in the different nuclei was analyzed where the general features
are understood from the contribution of the different orbitals
which become occupied. Especially, it was concluded that the
specific features of the yrast ultrahigh-spin band in   
$^{158}$Er are best understood
if this band is built in the TSD minimum with $\gamma < 0$,
i.e. for rotation around the intermediate axis. 
\bigskip

This work was supported in part by the Swedish Research Council and the Iranian Ministry of Science, Research and Technology.

\end{document}